\documentclass[a4paper,fleqn,usenatbib]{mnras}

\usepackage{mathptmx}
\usepackage{txfonts}

\usepackage[T1]{fontenc}
\usepackage{ae,aecompl}

\usepackage{array}
\usepackage{graphicx}
\usepackage{natbib}
\usepackage{lscape}
\usepackage{rotating}

\newcommand{\kms}{km\,s$^{-1}$}

\title[How unique is Plaskett's star?]{How unique is Plaskett's star? A search for organized magnetic fields in short period, interacting or post-interaction massive binary systems\thanks{Based on observations collected at the European Organisation for Astronomical Research in the Southern Hemisphere under ESO programme 095.D-0075.}}

\author[ Y. Naz\'e et al. ]{Ya\"el Naz\'e$^{1}$\thanks{FNRS Research Associate, E-mail: naze@astro.ulg.ac.be}, Coralie Neiner$^2$, Jason Grunhut$^{3,4}$, Stefano Bagnulo$^5$, Evelyne Alecian$^6$, \newauthor Gregor Rauw$^1$, Gregg A. Wade$^7$, and the BinaMIcS collaboration
\\
$^1$ Groupe d'Astrophysique des Hautes Energies - STAR, Institut d'Astrophysique et de G\'eophysique - B5c, Universit\'e de Li\`ege, 19c \\ All\'ee du 6 Ao\^ut, B-4000 Sart Tilman, Belgium\\
$^2$ LESIA, Observatoire de Paris, PSL Research University, CNRS, Sorbonne Universit\'es, UPMC Univ. Paris 06, Univ. Paris Diderot, \\ Sorbonne Paris Cit\'e, 5 place Jules Janssen, 92195 Meudon, France\\
$^3$ European Southern Observatory (ESO), Karl Schwarzschild Strasse, 85 748, Garching bei M\"unchen, Germany\\
$^4$ Dunlap Institute for Astronomy and Astrophysics, University of Toronto, Rm 101, 50 St. George Street, Toronto, ON, M5S 3H4, Canada\\
$^5$ Armagh Observatory and Planetarium, College Hill, Armagh, BT61 9DG, UK\\
$^6$ Universit\'e Grenoble Alpes, IPAG, CNRS, 38000 Grenoble, France \\
$^7$ Dept. of Physics, Royal Military College of Canada, P.O. Box 17000, Station Forces, Kingston, ON, Canada, K7K 4B4
}

\pubyear{2017}

\begin{document}
\label{firstpage}
\pagerange{\pageref{firstpage}--\pageref{lastpage}}
\maketitle

\begin{abstract}
Amongst O-type stars with detected magnetic fields, the fast rotator in the close binary called Plaskett's star shows a variety of unusual properties. Since strong binary interactions are believed to have occurred in this system, one may wonder about their potential role in generating magnetic fields. Stokes $V$ spectra collected with the low-resolution FORS2 and high-resolution ESPaDOnS and Narval spectropolarimeters were therefore used to search for magnetic fields in 15 interacting or post-interaction massive binaries. No magnetic field was detected in any of them, with 0\,G always being within 2$\sigma$ of the derived values. For 17 out of 25 stars in the systems observed at high-resolution, the 90\% upper limit on the individual dipolar fields is below the dipolar field strength of Plaskett's secondary; a similar result is found for five out of six systems observed at low resolution. If our sample is considered to form a group of stars sharing similar magnetic properties, a global statistical analysis results in a stringent upper limit of $\sim200$\,G on the dipolar field strength. Moreover, the magnetic incidence rate in the full sample of interacting or post-interaction systems (our targets + Plaskett's star) is compatible with that measured from large surveys, showing that they are not significantly different from the general O-star population. These results suggest that binary interactions play no systematic role in the magnetism of such massive systems.
\end{abstract}




\section{Introduction}
Prior to the early 2000s, no direct detection of magnetic fields in O-type stars had been reported. However, such fields were strongly suspected to exist based on several indirect pieces of evidence, such as the presence of strong magnetic fields in the remnants of massive stars (pulsars and magnetars), of synchrotron radio emission in some massive binaries \citep[for a review on non-thermal emitters, see][]{ben10}, or of particular phenomena such as periodic modulations of line profiles (as e.g. in $\theta^1$\,Ori\,C, \citealt{sta96}). Direct evidence of magnetism was finally acquired in the beginning of the 21st century. $\theta^1$\,Ori\,C was the first O-type star found to be magnetic \citep{don02}, HD\,191612, an Of?p star, was the second \citep{don06}, and many more have followed since then. Currently, about a dozen O-stars and several tens of early B-stars are known to be magnetic \citep[see e.g. a summary by][]{pet13}, leading to an inferred incidence rate of such fields of about 6--8\% amongst massive stars \citep{fos15,gru16}. 

Some of the first detections were made in binaries: for example, $\theta^1$\,Ori\,C and HD\,191612 are known to have companions. However, the orbital periods of these systems are long, implying that the binarity has little influence on the stellar properties. The case of Plaskett's star is very different: this ``star'' actually is a binary system harbouring two late-type O-stars in a rather tight orbit with a period of 14.4\,d \citep[as first demonstrated by][]{pla22}. It displays several peculiarities. The secondary star appears slightly more massive than the primary and rotates more than four times faster than its companion \citep{bag92,lin08}. The stars were classified O7.5I+O6I by \citet{bag92} and O8 III/I+O7.5 III by \citet{lin08} - but the latter authors noted that the secondary may actually be a O7V star that shows an apparent O7.5III spectrum because of its rapid rotation. A mismatch then exists between the spectroscopic masses and the dynamical ones, the latter being too large \citep{lin08,gru13}. Analyzing the optical spectra with atmosphere models further revealed that both stars are enriched in helium, that the secondary surface is depleted in nitrogen, and that the nitrogen abundance of the primary star is about 16 times that of the Sun while its carbon abundance is depleted by a factor of five  \citep[the N enrichment was already detected in X-ray data by \citealt{lin06}]{lin08}. In order to explain the unusual properties of the components, \citet{bag92} and \citet{lin08} proposed that the system is a post mass transfer binary in which the (currently) rapidly rotating secondary star received mass and angular momentum from its companion. In parallel, the Doppler maps derived from a tomographic analysis of the H$\alpha$ and He\,{\sc ii}$\lambda$\,4686 lines revealed an annular emission region which was interpreted in terms of a flattening of the secondary's wind \citep{lin08}. This was subsequently understood as a region of magnetically confined winds when a strong magnetic field was detected on this star \citep[$B_d>2.85$\,kG, ][]{gru13}. It is worth noting that Plaskett's secondary is the only known rapidly-rotating magnetic O-star, hence the only O-star to appear in the centrifugally-supported magnetosphere zone of the confinement-rotation diagram \citep{pet13}. 

The peculiar properties and evolutionary history of this system lead to speculations regarding the origin of the magnetic field of Plaskett's secondary, with two main possibilities. On one hand, Plaskett's secondary is a massive star, and global (i.e. not localized) magnetic fields in massive stars are generally thought to be fossil \citep[e.g.][ for a review see \citealt{nei15iaus} and references therein]{mos01}. This could be the case of Plaskett's secondary and, in this case, it would simply be amongst the few percent of magnetic objects in the massive star population. Its peculiarities - linked to binary interactions (stellar tides, see \citealt{pal14}, and previous mass transfer event, \citealt{lin08}) - would then be a coincidence. On the other hand, binarity could have played a key role, being directly responsible for the generation of a global magnetic field. This could occur through the triggering of differential rotation \citep[e.g.][]{spr02,bra06}. Such a scenario was proposed by several authors, in particular to explain magnetism in massive stars as a result of stellar mergers \citep{fer09,lan14}. Magnetism acquired after a merging event would naturally concern (currently) single magnetic objects or magnetic stars in long-period systems, but this idea cannot be applied directly to Plaskett's star as it is a short-period system. However, ``almost-merging'' systems, after a common-envelope phase or a mass-transfer episode, should also experience strong shear, though discussions arose on whether such processes would also constitute a viable road towards magnetism. Indeed, some Be stars in binary systems may be the products of a past mass-transfer \citep[and references therein]{van97,swa05,dem13} but no large-scale magnetic field was detected for those objects \citep[see \citealt{wad16be} and Sect. 3.3 of][and references therein]{riv13}. Nevertheless, Plaskett's star clearly appears as a potential candidate for this scenario \citep[e.g.][]{lan14,sch16}. If true, it would have important implications for understanding the general pathways leading to the production of magnetic fields in massive stars.

In this study we set out to directly assess the validity of this scenario, which has never before been done, by verifying the magnetic status of ``twins'' of Plaskett's star, defined as short-period massive binaries that interact or have interacted in the recent past. If binary interactions, in particular mass transfer events, constitute a dominant process for the generation of magnetic fields of massive stars, then most (if not all) of these systems would possess a magnetic component. In that case, we would thus have identified a specific group of magnetic O-stars, the second such category after the group of Of?p stars which were all found to be magnetic since the pioneering detection of 2006 (HD\,191612, \citealt{don06,wad11}; HD\,148937, \citealt{hub08,wad12}; HD\,108, \citealt{mar10}; NGC1624-2, \citealt{wad12dash}; CPD\,$-28^{\circ}$2561, \citealt{hub12,wad15}). On the contrary, if mass transfer events are not a source of strong, large-scale magnetic fields, then the magnetic incidence in these systems would not differ significantly from that found for large stellar samples \citep{fos15,gru16}. In this paper, we report the results of such a campaign. Sections 2 and 3 present the targets and the observations used in this study, respectively, while Sect. 4 provides the results and Sect. 5 reports our conclusions.

\section{The sample}
The first step of the project was to select interacting or post-interaction massive binaries. The key features to identify such systems are the sizes of the stars (e.g. filling their Roche lobes or close to doing so), but also the detection of non-solar abundances and/or of rotational asynchronicity, as well as a mismatch between predictions from evolutionary models of isolated stars and the observed stellar properties (age, masses, abundances, and luminosities). In addition, these binaries need to be bright enough to be observable with the current generation of spectropolarimeters, but they also ought to be well known, i.e. to have reliable physical parameters, so that there is no ambiguity on their evolutionary status. \citet{gie03} provided lists of well-studied double-lined spectroscopic (SB2) binary systems located in the Galaxy and the Magellanic Clouds, among which he separately listed unevolved cases (his Table 1) and evolved cases (his Table 2), the latter ones being subdivided between systems before contact, in contact, and after mass transfer. \citet{pen08} also listed binaries in which one star fills its Roche lobe, complementing the list. Eliminating the complex high-multiplicity systems $\delta$\,Ori and Cyg\,OB2\,5, which could blur the picture, while adding HD\,149404 and LSS\,3074, two other post-Roche Lobe Overflow (RLOF) systems not listed in these references but whose interactions were recently analyzed (see details below), we ended up with 15 relatively bright targets:

\begin{table*}
  \caption{ Summary of the properties of our targets and of Plaskett's star }
  \label{prop}
  \begin{tabular}{lcccccccc}
  \hline
ID & $V$& Sp. Types & P (d) & e & C/SD? & AA? & Mass & $v\sin(i)$ (\kms) \\
   &    &           &       &   &       &     & gainer  & prim-sec-tert \\
\hline
Plaskett's star & 6.06& O8III/I+O7.5III& 14.4 & 0 & N & Y& sec. & 77-370$^*$ \\ 
HD\,1337  & 6.14& O9.7III+O9.5V      & 3.5 & 0 & Y &  & sec. & 118-82$^*$ \\
HD\,25638 & 6.93& O9.5V-9IV+B0-0.5+B0III-III & 2.7 & 0 & N &  & prim. & 137-69-40$^*$ \\
HD\,35652 & 8.39& O9.5V+B0.5IV-V+B   & 1.8 & 0 & Y &  & prim. & 172-159-71$^*$ \\
HD\,35921 & 6.85& O9III+O9.5III+B    & 4.0 & 0 & Y &  & prim.? & 202-122-27$^*$ \\
HD\,57060 & 4.95& O8.5I+O9.7V        & 4.4 &0.1& Y &  & sec. & 107-193$^*$ \\
HD\,100213& 8.41& O7.5V+O9.5V        & 1.4 & 0 & Y &  &  & 222-191 \\
HD\,106871& 8.48& O8V+B0.5:          & 3.4 & 0 & Y & Y& prim. & \\
HD\,115071& 7.97& O9.5V+B0.2III      & 2.7 & 0 & Y &  & prim. & 101-132 \\
HD\,149404& 5.47& O7.5If+ON9.7I      & 9.8 & 0 & N & Y& prim. & 56-80$^*$ \\
HD\,152248& 6.05& O7.5III(f)+O7III(f)& 5.8 &0.1& N &  &  & 139-128$^*$ \\
HD\,190967& 8.16& O9.5V+B1I-II       & 6.5 & 0 & Y &  & prim. & 230-115$^*$ \\
HD\,209481& 5.55& O9III+ON9.7V       & 3.1 & 0 & N & Y& prim. & 127-93$^*$ \\
HD\,228854& 8.65& O7.3V+O7.7V        & 1.9 & 0 & Y &  &  & 223-203$^*$ \\
LSS\,3074 & 11.7& O4If+O7.5          & 2.2 & 0 & N & Y& sec. & 110-127 \\
XZ\,Cep   & 8.51& O9.5V+B1III        & 5.1 & 0 & Y &  & prim. & 104-172$^*$ \\
\hline
\end{tabular}

{\footnotesize Evidence for current or past interaction (for references, see text): ``C/SD'' = in a contact or semi-detached configuration, ``AA'' = abundances anomalies detected - note that there are also other anomalies detected (e.g. too low masses, asynchronous rotation,...), see text for details. The projected rotational velocities $v\sin(i)$ are given for primary and then secondary then tertiary, they come from our analysis of high-resolution spectra when an asterisk is added and from literature otherwise (\citealt{how97} for HD\,100213, otherwise see dedicated items in text for literature references). }
\end{table*}

\begin{itemize}
\item {\it HD1337$=$AO\,Cas} (O9.7III+O9.5V) consists of two stars in a circular orbit of 3.5\,d period \citep{sti97,lin08th}. The analysis of the eclipses yields a semi-detached configuration, with a `hot spot' on the secondary possibly corresponding to the impact of an accreting stream \citep{lin08th}. Furthermore, stellar masses and radii appear too small compared to typical values for massive stars of the same spectral types, reinforcing the mass-transfer scenario \citep{lin08th}.
\item {\it HD25638$=$SZ\,Cam} (O9.5V-9IV+B0-0.5V) comprises an eclipsing, circular SB2 system with a period of 2.7\,d \citep[and references therein]{gor15}. The derived radius of the primary is too large for a main sequence star while the mass of the secondary is too small for its spectral type \citep{lor98}. Furthermore, the derived masses, luminosities, and temperatures could not be fitted by theoretical evolutionary tracks, ruling out an `interaction-free evolution' \citep{har98,lor98,tam12}. 
\item {\it HD35652$=$IU\,Aur} (O9.5V+B0.5IV-V) comprises an eclipsing, semi-detached binary with a circular orbit and a period of P=1.8\,d \citep[and references therein]{har98,ozd03}. The derived masses and gravities of the stars cannot be explained by the separate evolution of massive stars born at the same time, suggesting a past mass-transfer event \citep[see in particular their Fig. 9]{har98}.
\item {\it HD35921$=$LY\,Aur} (O9III+O9.5III) is a circular binary with a period of 4.0\,d \citep[and references therein]{sti94,zha14}. The analysis of the eclipses suggests a semi-detached or slightly overcontact configuration \citep{li85,dre89,may13} and the derived masses appear slightly too low \citep{sti94,may13}, favouring an evolutionary scenario where mass-transfer has occurred.
\item {\it HD57060$=$29\,CMa} (O8.5I+O9.7V) is a slightly eccentric binary with a period of 4.4\,d \citep{bag94,sti97,lin08th}. The photometric eclipses suggest some interaction between the stars (from semi-detached to overcontact configuration, \citealt{bag94,lin08th,ant11}) and the anomalous luminosities also seem to indicate an on-going mass-transfer. Furthermore, a tomographic analysis of the H$\alpha$ and He\,{\sc ii}\,4686 lines  \citep{lin08th} and discrepancies in the mass ratio determination \citep{ant11} suggest a flattened, disk-like secondary wind, similar to that found in Plaskett's star. Finally, a wind-wind collision is also present in the system \citep{lin08th}.
\item {\it HD100213$=$TU\,Mus} (O7.5V+O9.5V) is a binary with $e=0$ and $P=1.4$\,d \citep{lin07,pen08}. Mutual irradiation of the stars in this very short-period system leads to the observation of line strength variations along the orbit \citep{lin07,pal13}. The eclipses of the system indicate that the stars are in contact while the masses derived from the orbital solution are lower than expected from evolutionary models, suggesting a past episode of mass-transfer \citep[and references therein]{lin07,qia07,pen08}. 
\item {\it HD106871$=$AB Cru} (O8V+B0.5:) appears as a semi-detached, eclipsing binary with a circular orbit and a period of 3.4\,d \citep{lor94}. A consistent fit to the derived properties of the stars (effective temperatures, radii, masses, luminosities) cannot be found with evolutionary tracks considering the absence of interaction while anomalous hydrogen and helium abundances are detected, both facts suggesting the system to have undergone mass-transfer \citep{lor94}.
\item {\it HD115071} (O9.5V+B0.2III) consists of a semi-detached binary with a circular orbit and a period of 2.7\,d \citep{pen02}. The ellipsoidal variations of its lightcurve indicate that the secondary star fills its Roche lobe. The derived masses are smaller than expected, suggesting mass-transfer to have occurred in a very recent past \citep[and references therein]{pen02}. 
\item {\it HD149404} (O7.5If+ON9.7I) is a non-eclipsing, detached binary with a circular orbit and a period of 9.8d \citep{rau01,tha01}. Emission lines such as H$\alpha$ and He\,{\sc ii}\,4686 display phase-locked variations indicating the presence of a wind-wind collision  \citep{rau01,tha01}. In addition, the secondary star is of the ON type, i.e. it presents anomalous abundances which hint at a past interaction \citep{rau01,tha01}. A recent, more detailed study further reveals an asynchronous rotation of the two stars, a significant nitrogen overabundance of the secondary star, and peculiar positions of the stars in the HR diagram \citep{rau16}. All of these properties clearly disagree with expectations of single star evolutionary models, yielding further support to the past mass-exchange scenario \citep{rau16}. 
\item {\it HD152248} (O7.5III(f)+O7III(f)) is a slightly eccentric and eclipsing binary with a period of 5.8\,d \citep{san01,may08}. Phase-locked variations in H$\alpha$ and He\,{\sc ii}\,4686 lines as well as in the X-ray emission arise from a wind-wind collision located between the two stars \citep{san01,san04}. The derived masses are lower than expected from evolutionary tracks and both components appear close to filling their Roche lobes, suggesting a past mass-transfer episode \citep{san01}. Note that tidal effects are reinforced by a small, non-zero, eccentricity.
\item {\it HD190967} (O9.5V+B1I-II) consists of an eclipsing binary in a semi-detached configuration and with a period of 6.5\,d and $e=0$ \citep{har97,kum05}. In a mass-gravity diagram, the derived values are at odds with predictions of evolutionary models for stars of the same age, indicating a past episode of rapid mass-transfer \citep{har97,kum05}. Note that, to best fit the lightcurve, either bright spots or an accretion disk are needed \citep{kum05,dju09}.
\item {\it HD209481$=$LZ\,Cep} (O9III+ON9.7V) is a binary with a circular orbit and a period of 3.1\,d. The ellipsoidal variations of its optical lightcurve indicate that both stars almost fill their Roche lobes \citep{mah11,pal13b}. Several lines of evidence point towards a past mass-transfer episode \citep{har98,mah11}: the photosphere of the secondary star appears strongly enriched in helium and nitrogen but depleted in carbon and oxygen, there is a slight asynchronicity between the rotation periods, and the evolutionary masses are higher than the masses inferred from the orbital solution. 
\item {\it HD228854} (O7.3V+O7.7V) is a system with a circular orbit and a period of 1.9\,d \citep[and references therein]{yas13}. The analysis of the eclipses indicates filling of the Roche lobe(s) (either overcontact, \citealt{deg99,qia07}, or contact/semi-detached, \citealt{may02,yas13}) and the derived physical properties (in particular masses and gravities) disagree with theoretical expectations for (single) massive star evolution \citep{har97}, pointing towards a post mass-transfer situation.
\item {\it LSS3074} (O4If+O7.5f) is a circular binary with a period of 2.2\,d \citep[and Raucq et al. 2017 submitted]{nie92,haf94}. H$\alpha$ and He\,{\sc ii}\,4686 lines present line profile modulations that are reminiscent of those observed in HD\,149404, suggesting wind-wind collision to be present (Raucq et al., 2017, submitted). Photometric, spectroscopic, and polarization measurements led to the derivation of the physical parameters for the system \citep[and Raucq et al. 2017 submitted]{nie92}. The anomalous abundances as well as the mismatch between evolutionary tracks and observed masses and luminosities suggest a mass-transfer in the system (Raucq et al. 2017 submitted).
\item {\it XZ\,Cep} (O9.5V+B1III) is a non-eccentric, eclipsing system with a period of 5.1\,d \citep{har97}. The secondary star fills its Roche lobe while the derived stellar properties are at odds with predictions from evolutionary models (incompatible ages for the derived masses and gravities, \citealt{har97}), suggesting a previous episode of mass-transfer.
\end{itemize}
 
Evidence for past or current interactions (abundance anomalies, Roche lobe filling,...) in the chosen systems has just been mentioned above (see Table \ref{prop} for a summary), but the presence of such interactions does not necessarily imply that they were actually efficient in modifying the stellar structure. In particular, Plaskett's secondary displays a high rotation rate ($\sim$300\,\kms, \citealt{bag92,lin08,gru13}), demonstrating that momentum exchange was very efficient for that object. For the stars in our sample, projected rotational velocities are also rather high and generally amount to 100--200\,\kms, with components in four systems (HD\,35921, HD\,100213, HD\,190967, and HD\,228854) reaching even higher velocities ($v\sin(i)>$200\,\kms\, see Table \ref{prop}). This confirms that our sample is well suited to find out whether magnetic fields are triggered in massive binary interactions.

\section{Observations}
For the Northern targets, high-resolution spectropolarimetry was acquired with ESPaDOnS at CFHT \citep{don03} and Narval at TBL \citep{aur03}. For the Southern targets, low-resolution spectropolarimetry was acquired with FORS2 at ESO \citep{app92,sze98}. Two targets, HD\,152248 and HD\,149404, were observed at both low and high resolutions, as they are visible from Chile as well as from Hawaii. 

\subsection{Low-resolution FORS2 spectropolarimetry}

\begin{figure}
\includegraphics[width=9cm]{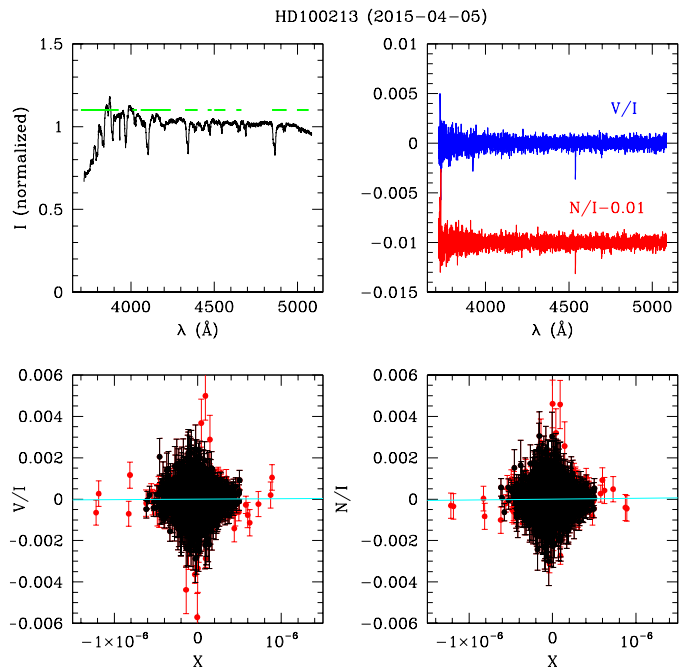}
\caption{{\it Top Left:} The spectrum of HD\,100213 taken on the second night of observation with FORS2, with the selected windows shown by a green thick line. {\it Top right:} Associated $V/I$ (top blue) and $N/I$ (bottom red) as a function of wavelength. {\it Bottom:} $V/I$ (left) and $N/I$ (right) as a function of $x_i$, with their best-fit straight lines shown in cyan. Rejected values, not considered for the fitting, appear in red. Note the null slopes of the best-fit straight lines, that indicate the absence of a significant magnetic field. }
\label{lowres}
\end{figure}

Low-resolution spectropolarimetric data of six targets were obtained with the Very Large Telescope equipped with FORS2 in Spring 2015 (ESO 095.D-0075, PI Naz\'e, see Table \ref{Bval}). These data were taken in service mode with the red CCD (a mosaic composed of two 2k$\times$4k MIT chips), no binning, a slit of 1'' and the 1200B grating ($R\sim 1400$). The observing sequence consisted of 8 subexposures with retarder waveplate positions of $+45^{\circ}$, $-45^{\circ}$, $-45^{\circ}$, $+45^{\circ}$, $+45^{\circ}$, $-45^{\circ}$, $-45^{\circ}$, $+45^{\circ}$. We reduced these spectropolarimetric data with IRAF as explained by \citet{naz12carina}: the aperture extraction radius was fixed to 20 pixels, the nearby sky background was subtracted, and wavelength calibration was performed from 3675 to 5128\AA\ (with pixels of 0.25\AA) considering arc lamp data taken at only one retarder waveplate position (in our case, $-45^\circ$). There was no indication of variation of the Stokes $I$ spectra between subexposures. We then constructed the normalized Stokes $V/I$ profile, as well as a diagnostic ``null'' profile $N/I$ \citep[see ][for details on the procedure and Fig. \ref{lowres} for an illustration in our case]{don97,bag09}. Finally, the associated longitudinal magnetic field was estimated by minimizing $\chi^2 = \sum_i \frac{(y_i - B_z\,x_i - a)^2}{\sigma^2_i}$ with $y_i$ either $V/I$ or the null profile at the wavelength $\lambda_i$ and $x_i = -g_\mathrm{eff}\ 4.67 \times 10^{-13} \ \lambda^2_i\ 1/I_i\ (\mathrm{d}I/\mathrm{d}\lambda)_i$ \citep{bag02}. This was done for $x_i$ in the interval between $-10^{-6}$ and $+10^{-6}$, after discarding edges and deviant points and after selecting spectral windows centered on stellar lines \citep[see][for further discussion]{naz12carina}. These windows comprise both primary and secondary absorption lines, but avoid emission lines (as some of them arise in wind-wind collision regions and are thus not representative of the stellar photosphere). The values reported in Table \ref{Bval} were obtained after rectifying the Stokes profiles, but similar values are found if no rectification is applied.

\begin{figure*}
\includegraphics[width=5cm]{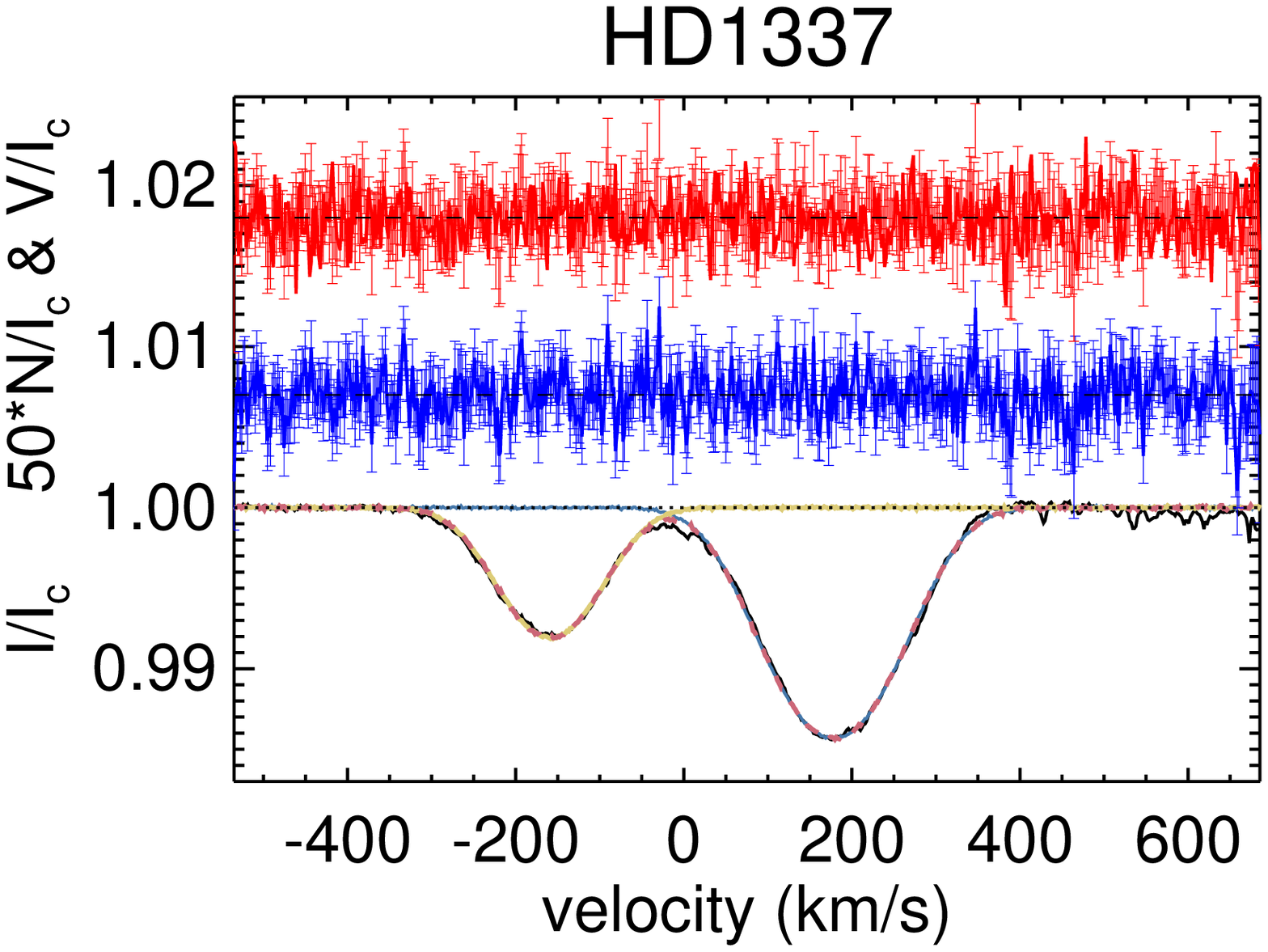}
\includegraphics[width=5cm]{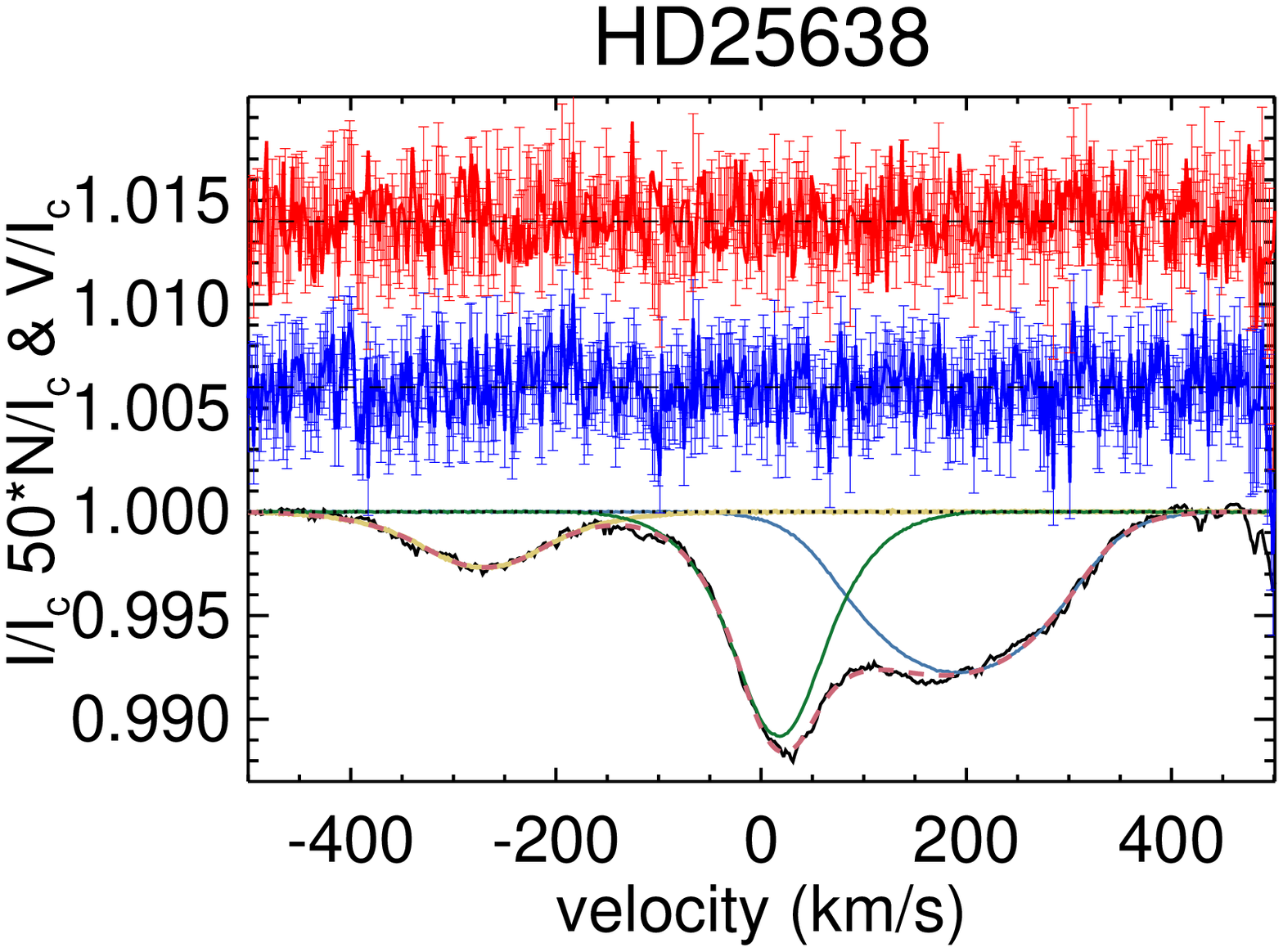}
\includegraphics[width=5cm]{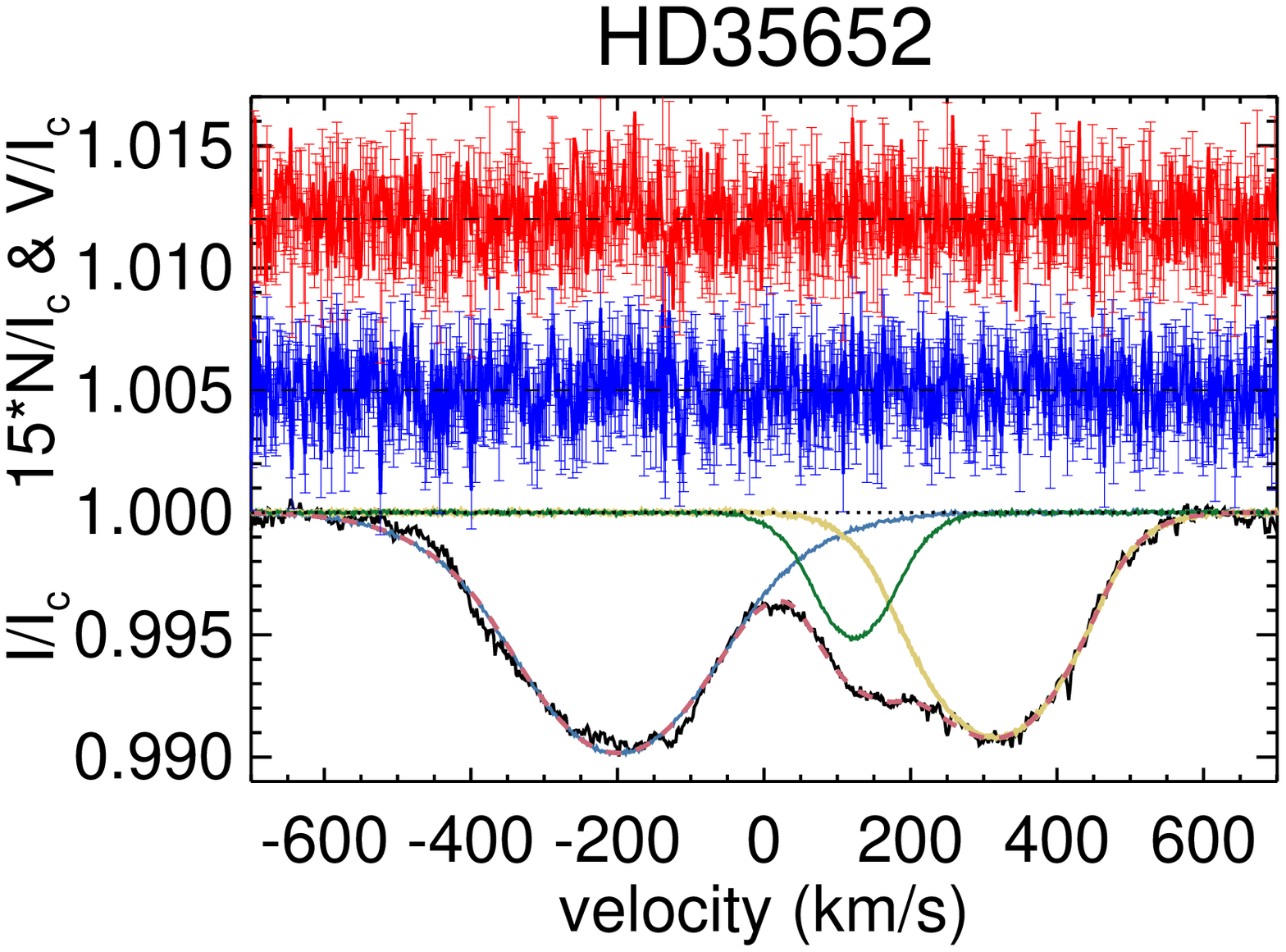}
\includegraphics[width=5cm]{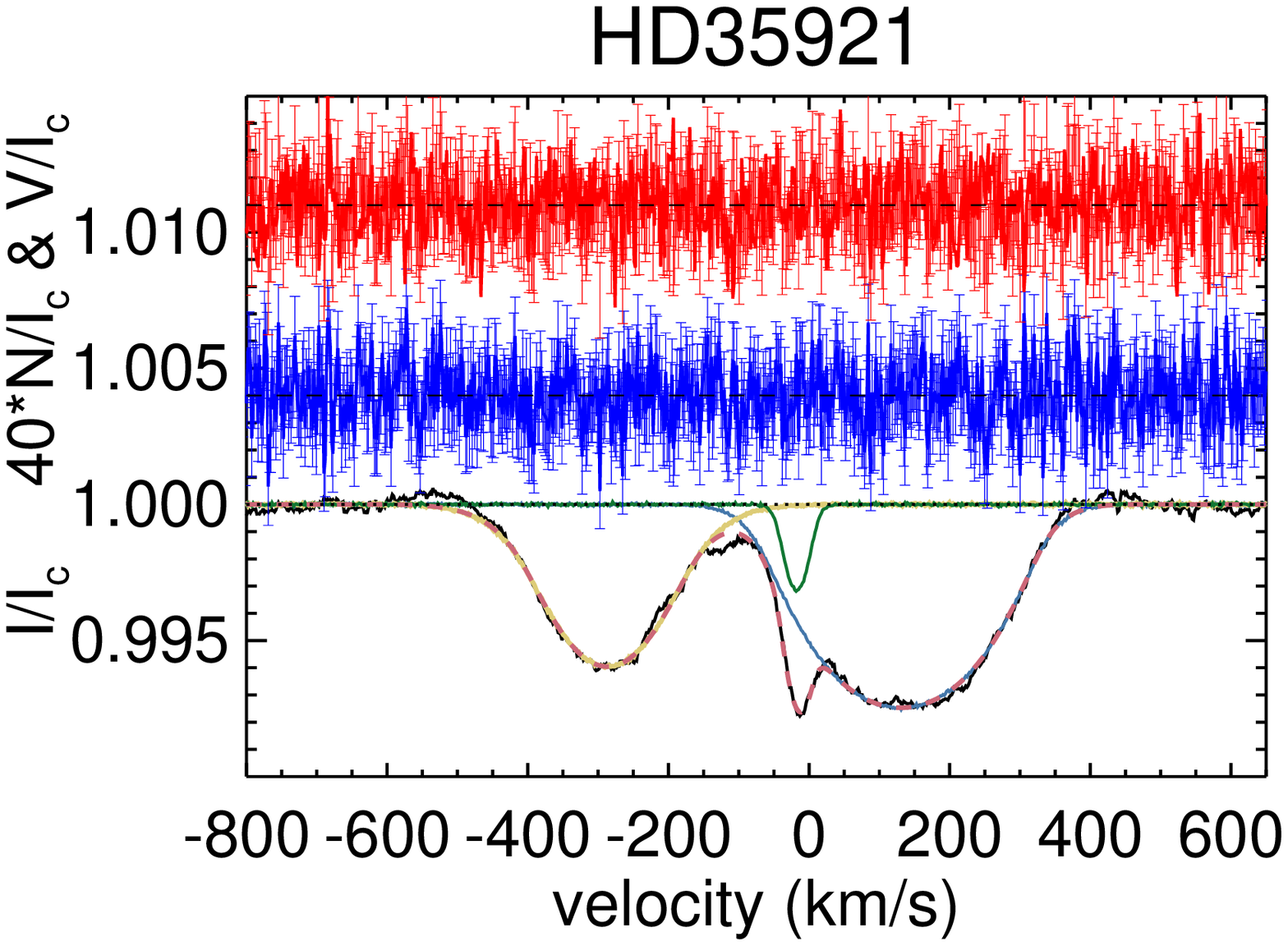}
\includegraphics[width=5cm]{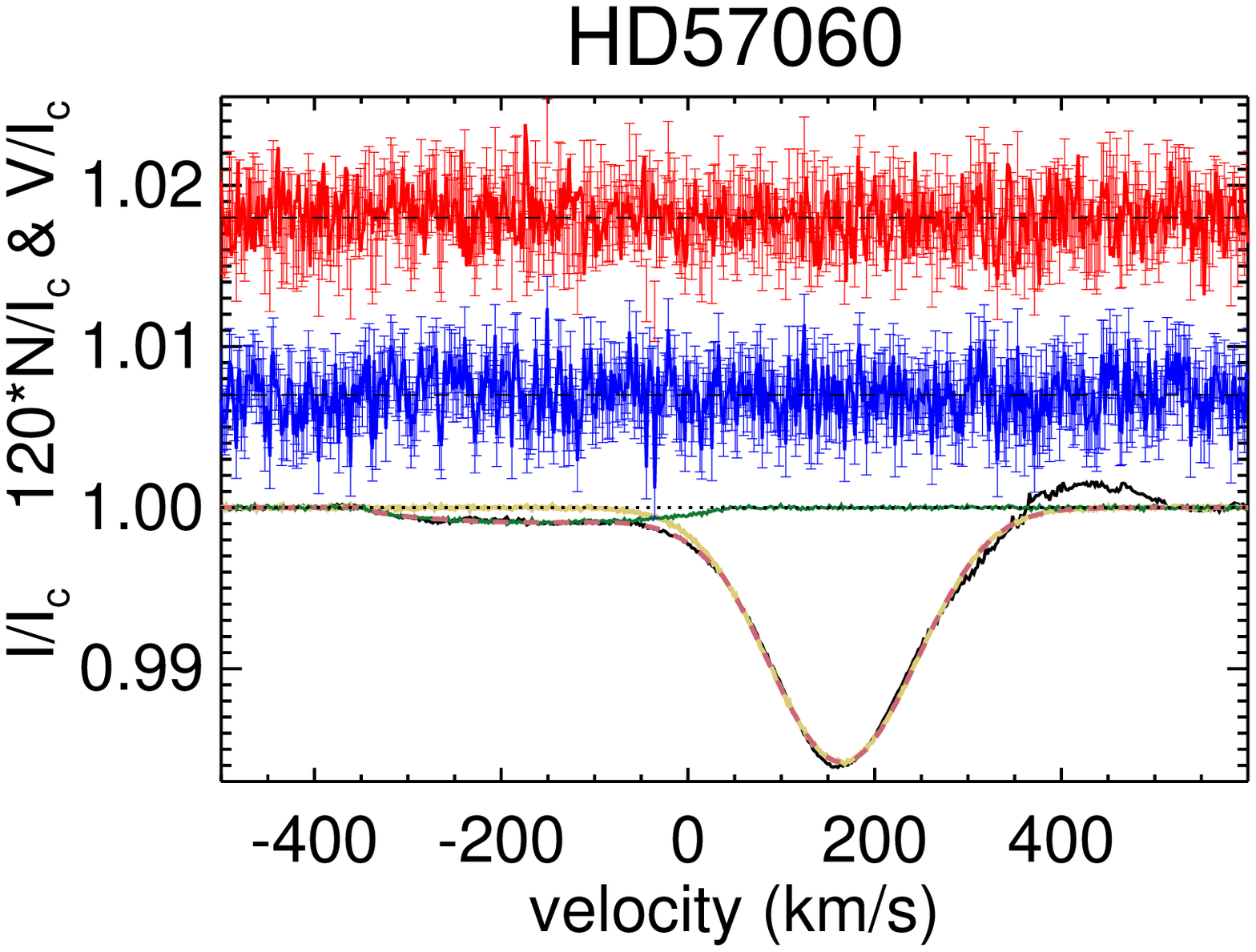}
\includegraphics[width=5cm]{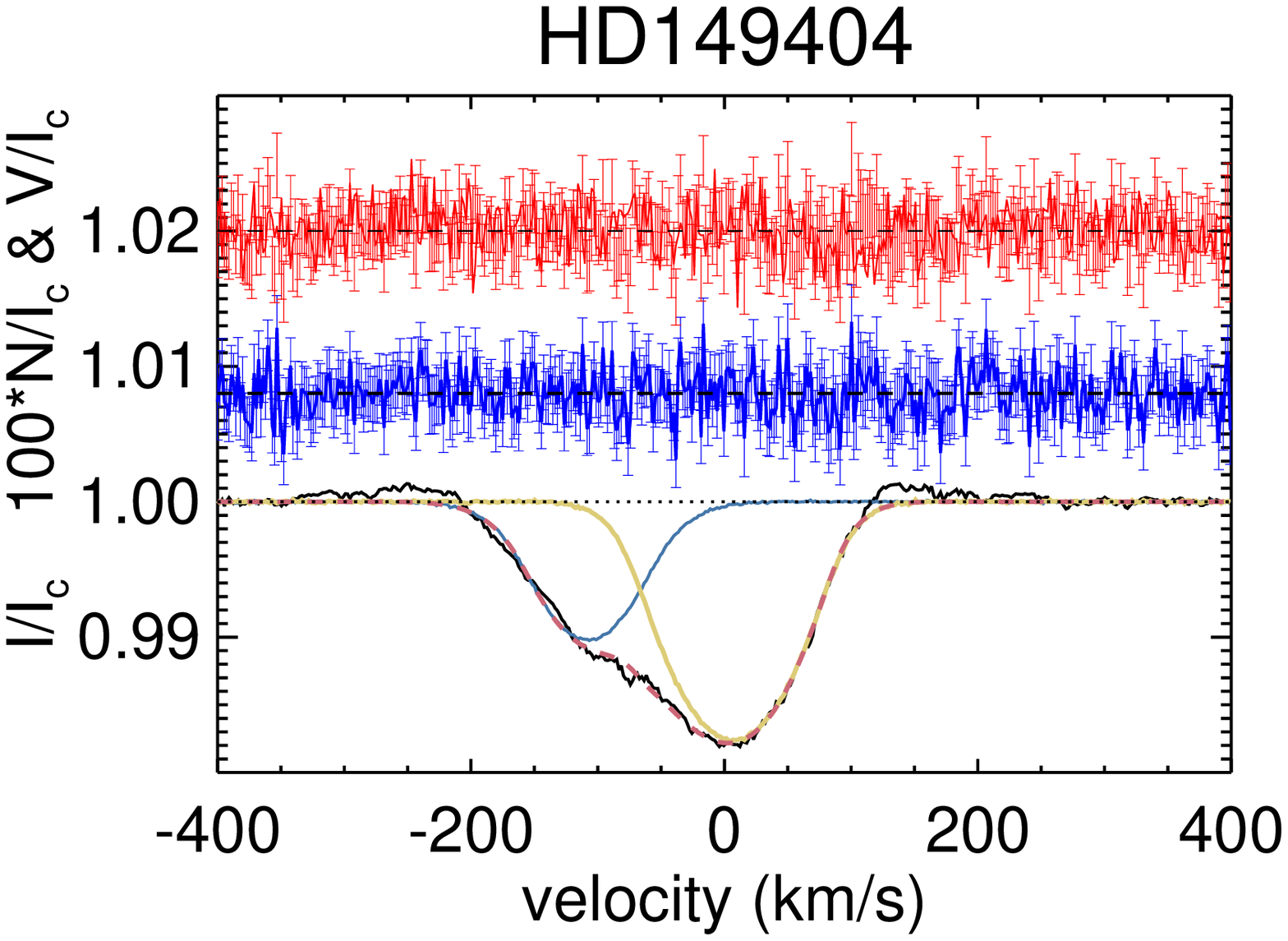}
\includegraphics[width=5cm]{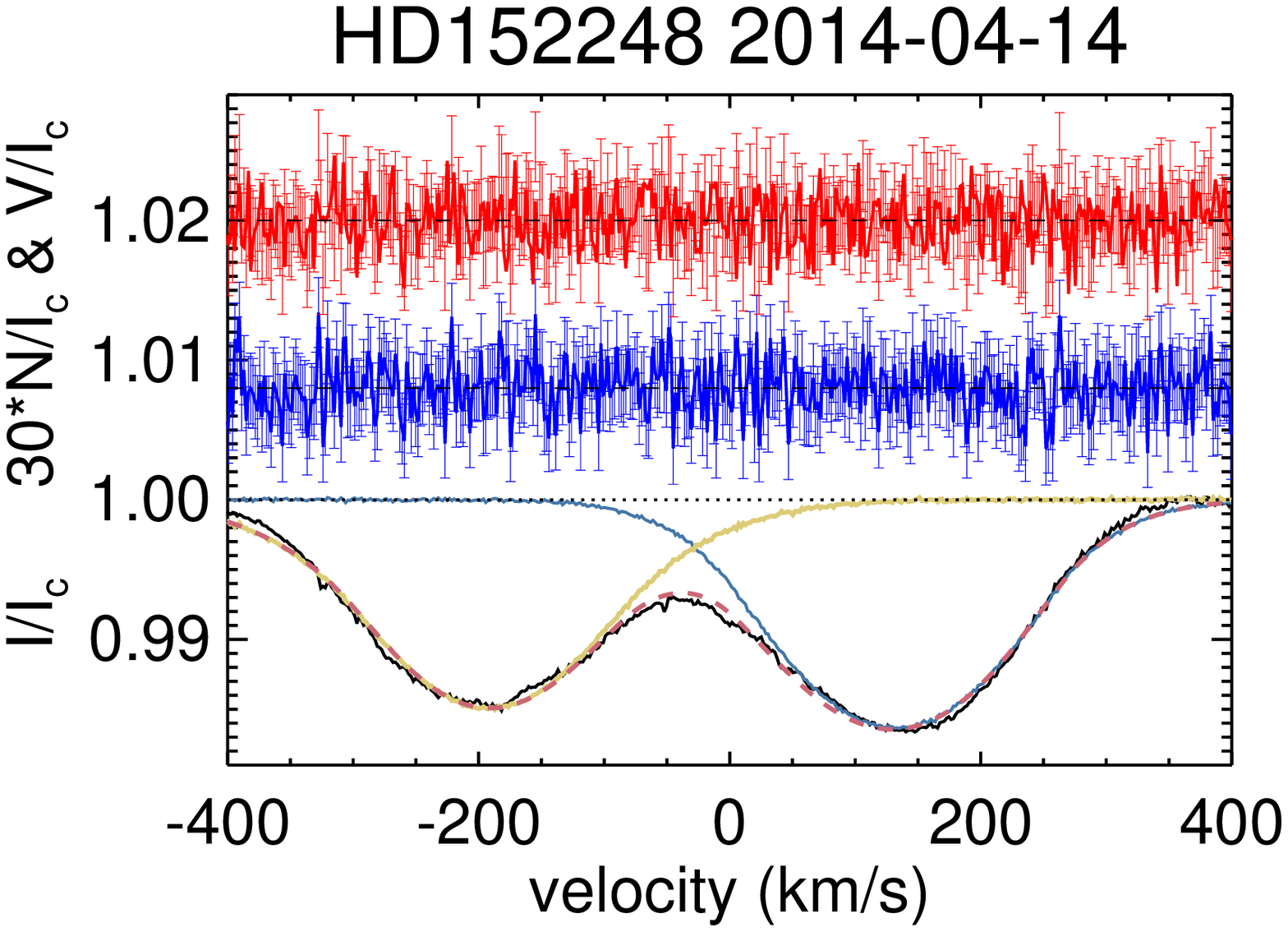}
\includegraphics[width=5cm]{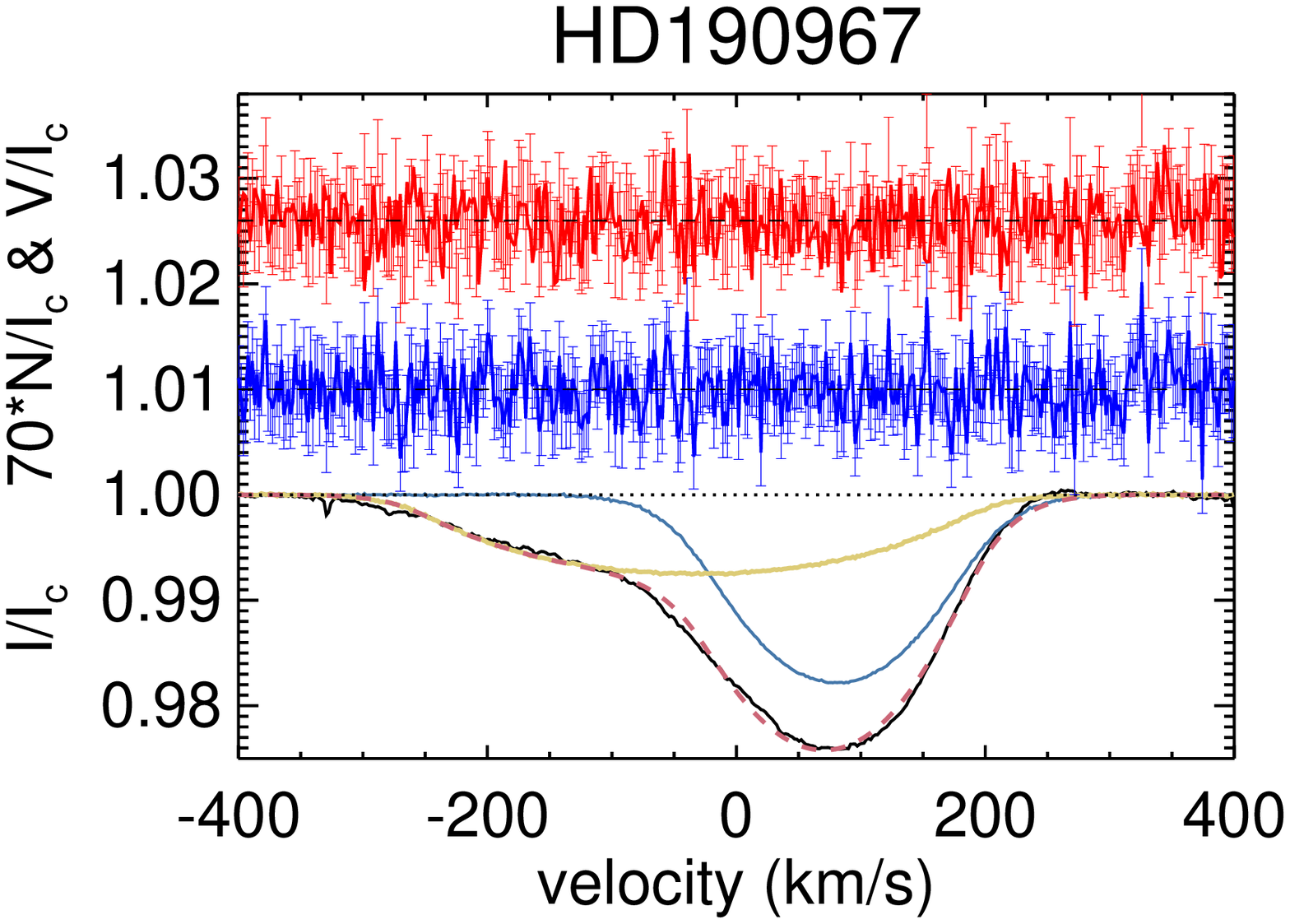}
\includegraphics[width=5cm]{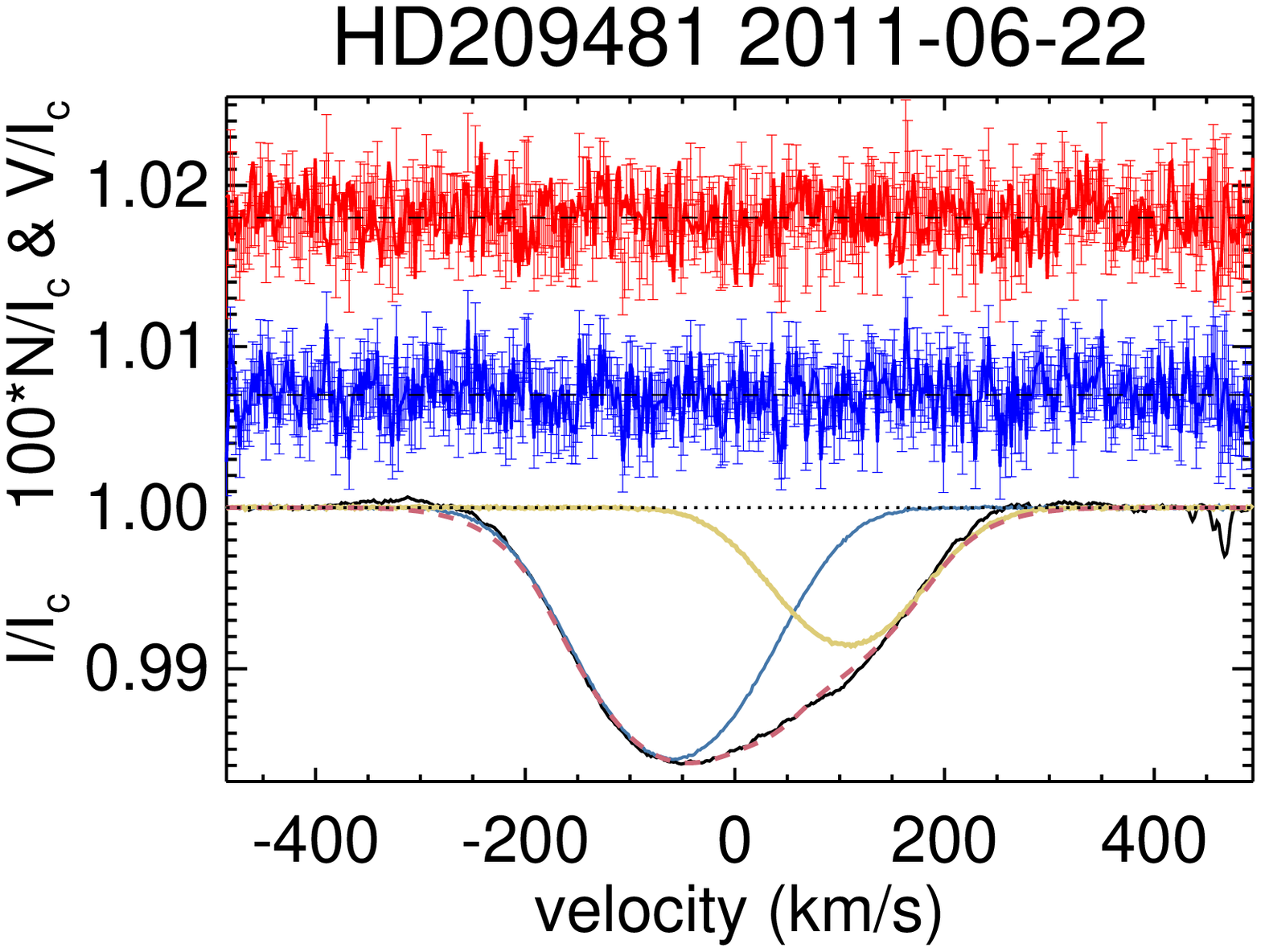}
\includegraphics[width=5cm]{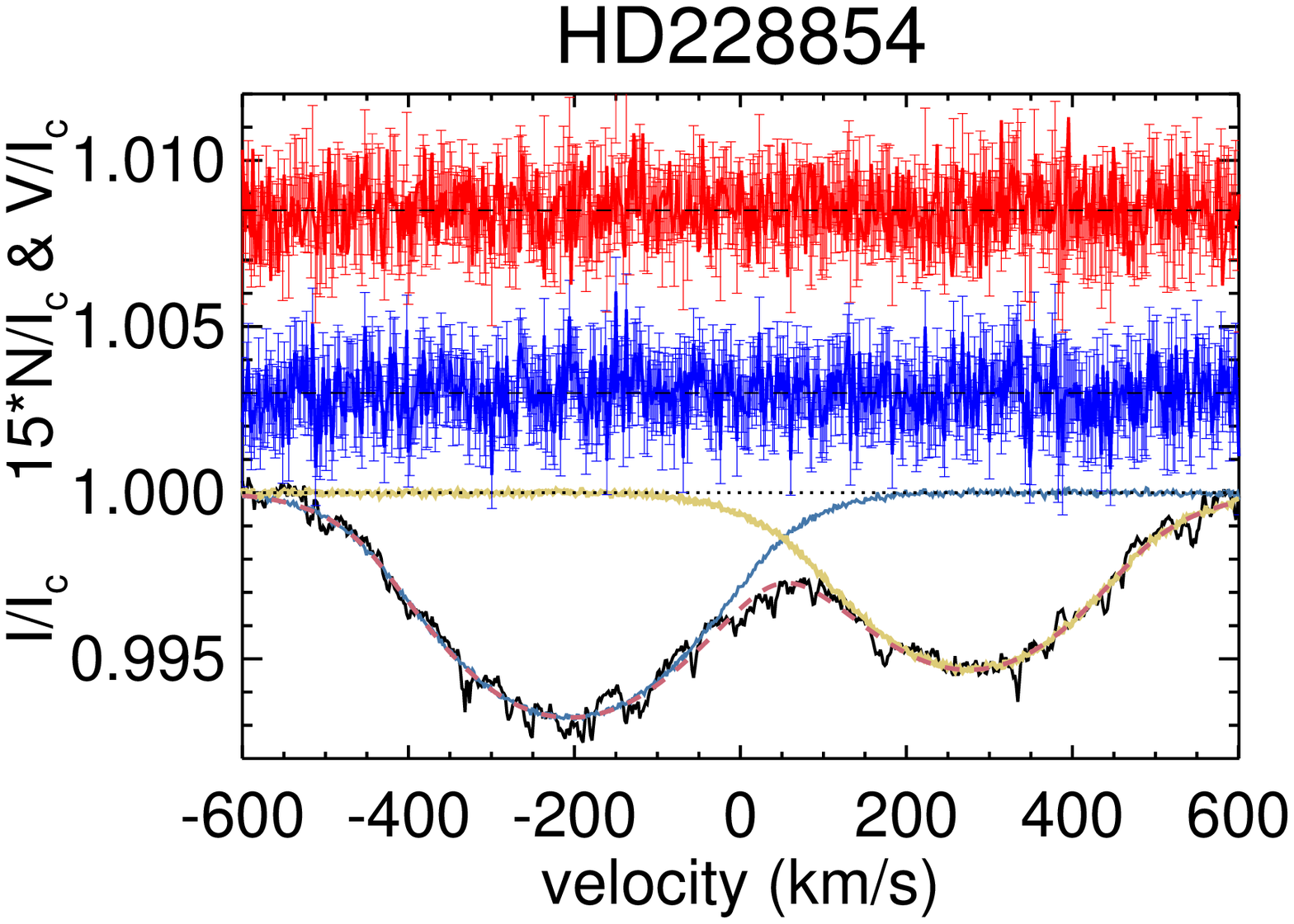}
\includegraphics[width=5cm]{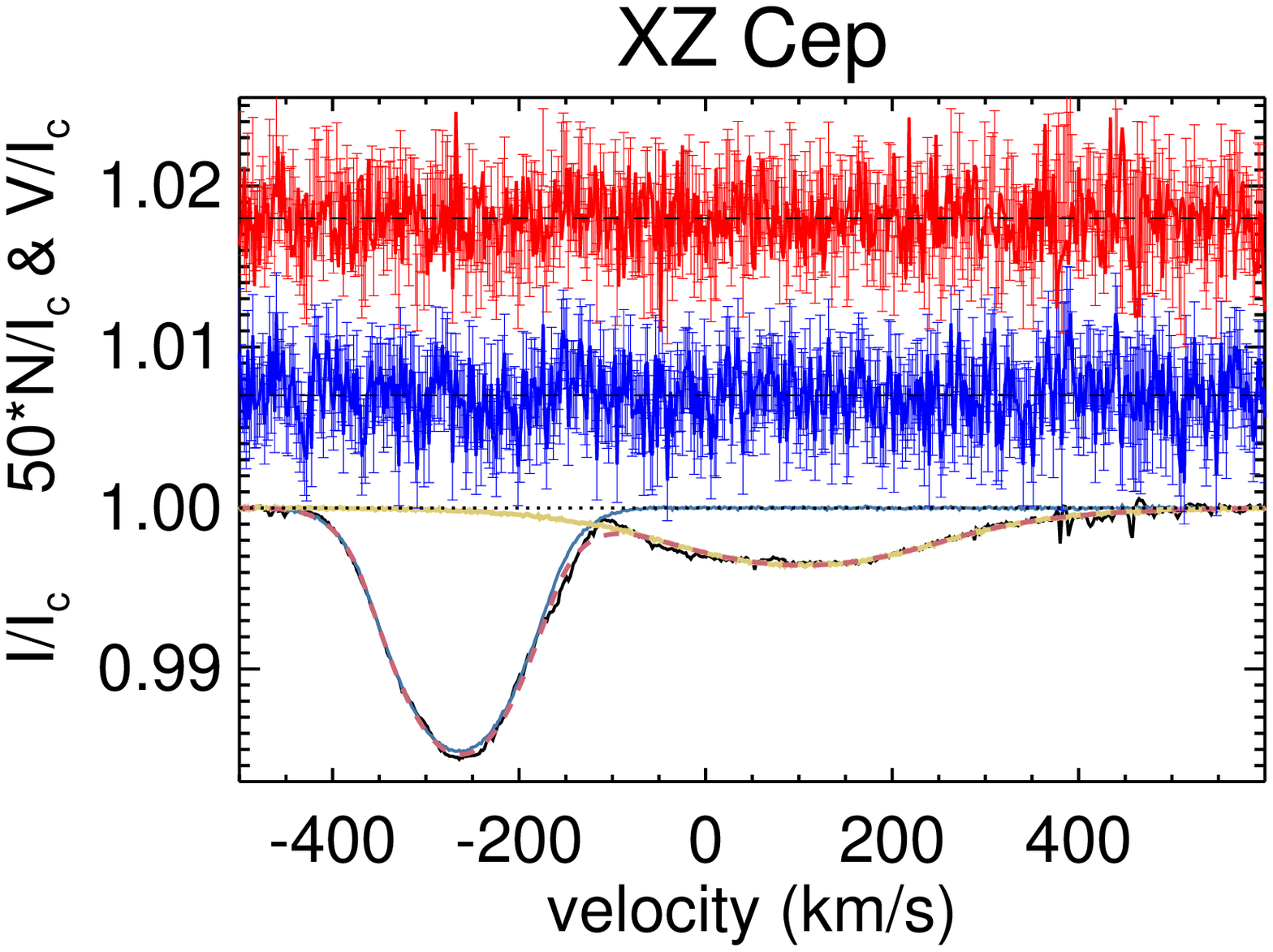}
\caption{LSD profiles derived for the BinaMIcS sample (coming from \citealt{gru16} for the three MiMeS targets - HD\,1337, HD\,35921, HD\,209481). Each panel provides, from top to bottom, the system name and the Stokes $V$, $N$, and $I$ profiles, respectively. The individual components of the $I$ profiles, derived from the disentangling procedure described in Sect. 3.2, are added with colours (yellow, blue, or green solid lines) along with their sum (dashed red lines). }
\label{highres}
\end{figure*}

\begin{table*}
  \caption{Results from the low-resolution spectropolarimetry obtained with FORS2.  }
  \label{Bval}
  \begin{tabular}{lccccccc}
  \hline
ID & $L_1/L_2$ & Exp. (s) &  SNR & Date & HJD-2450000. & $B_z$ (G) & $N_z$ (G) \\
\hline
HD\,100213 & $\sim2$ &8$\times$70 & 2470 & 2015-04-03 & 7115.687 & 34$\pm$48  &  2$\pm$47  \\ 
           &         &8$\times$70 & 2365 & 2015-04-05 & 7117.732 & 24$\pm$50  & 37$\pm$49  \\ 
HD\,106871 &$\sim2.5$&8$\times$80 & 2210 & 2015-04-03 & 7115.733 & $-31\pm$45 & 3$\pm$44   \\ 
           &         &8$\times$80 & 2380 & 2015-04-05 & 7117.754 & 32$\pm$39  & 60$\pm$38  \\ 
HD\,115071 &$\sim0.9$&8$\times$60 & 2350 & 2015-04-05 & 7117.837 & $-22\pm$34 & $-44\pm$33 \\ 
           &         &8$\times$60 & 2520 & 2015-05-01 & 7143.618 & $-42\pm$32 & 16$\pm$33  \\ 
HD\,149404 &$\sim0.7$&8$\times$7  & 1650 & 2015-05-16 & 7158.907 &214$\pm$119 &$-234\pm$117\\ 
           &         &8$\times$7  & 2130 & 2015-05-17 & 7159.585 & $-64\pm$80 & 3$\pm$69    \\
HD\,152248 & 0.8--1. &8$\times$10 & 2100 & 2015-05-02 & 7144.620 & $-25\pm$69 & 36$\pm$63   \\
           &         &8$\times$10 & 1845 & 2015-05-17 & 7159.570 & 29$\pm$59  & 90$\pm$59   \\
LSS\,3074  & 1.--2.5 &8$\times$536& 1400 & 2015-04-03 & 7115.809 &$-591\pm$619& 200$\pm$427 \\
           &         &8$\times$536& 1450 & 2015-04-05 & 7117.796 &$-452\pm$391& 187$\pm$362 \\
\hline
\end{tabular}

{\footnotesize From left to right, the columns show the binaries' names, the light ratios (see Sect. 2 for references), the exposure lengths, the peak signal-to-noise ratio in the 4000--5000\AA\ range, the observing times (both in YYYY-MM-DD format and in heliocentric Julian dates), and finally the derived magnetic field values and their associated errors for both the system's Stokes $V$ profile and from the null diagnostic profile. These $B_z$ and $N_z$ values were found using rectification and within selected spectral windows. }
\end{table*}

\subsection{High-resolution ESPaDOnS and Narval spectropolarimetry}

Most high-resolution spectropolarimetric data of our targets were collected in the context of the MiMeS \citep{wad16} and the BinaMIcS \citep{ale15} large programmes. They were acquired with ESPaDOnS, an echelle spectropolarimeter installed at the Canada-France-Hawaii Telescope. In addition, archival data from dedicated programmes (PIs Neiner, Bouret) were obtained for HD\,209481 in 2006 and 2009 using Narval, a similar spectropolarimeter installed at Pic du Midi (France). Biases, flat-fields, and ThAr calibrations were obtained at the beginning and at the end of each night. The data reduction was performed using Libre-Esprit, the dedicated reduction software based on Esprit \citep{don97}. There was no indication of variation of the Stokes $I$ spectra between subexposures. Normalization of the individual orders was then performed using low-degree polynomial fits. 

After the Stokes spectra were calculated, we followed the same procedure as \citet{gru16}. First, the Least-Squares Deconvolution technique (LSD, \citealt{don97,koc10}) was applied to all polarimetric spectra to increase the signal-to-noise, thereby helping to detect weak magnetic Zeeman signatures (see Fig. \ref{highres} for an illustration). We first constructed a line mask from the VALD2 database \citep{pis95,kup99} using the appropriate temperature and $\log(g)$ values for each star, assuming solar abundances and considering only lines deeper than 1\% of the continuum. We then excluded all hydrogen lines and lines blended with hydrogen. This yielded masks with between 430 and 1220 lines, depending on the star, over the 3700--9800\AA\ interval. The line depths were then adjusted to provide the best fit to the observed Stokes $I$ spectrum of each target. Note that regularisation \citep[see][]{koc10} and clipping of the deviant points were both applied to improve the final signal-to-noise of the LSD profiles. 

Since our targets are binaries, the second step was to derive the individual properties of each component (see Fig. \ref{highres} and \citealt{gru16} for details on the procedure). To this aim, multiple absorption components with profiles derived from the convolution of a rotationally broadened profile with a radial-tangential macroturbulence profile were fitted to the Stokes $I$ profiles. Since this disentangling process often leads to degenerate solutions, relevant information (e.g. $v \sin(i)$) from the literature was used to constrain the fits, whenever needed. The derived RVs were checked for compatibility with the known orbital solutions of these systems, i.e.  we verified that each observed pair of RVs agreed with the solution at {\it some} phase during the orbit (accurate orbital phases at current epoch cannot be computed due to the large accumulated phase uncertainties). Disentangled $I$ profiles were then obtained for each system component, and used to estimate individual $V$ and $N$ contributions. Note that, if line profiles overlap and one component is magnetic, this procedure does not account for the possible magnetic contamination caused by the Stokes $V$ signal of the companion. 

Finally, the magnetic properties were derived from the Stokes $V$ profiles. We calculated in each case the probability that deviations of the observed $V/I_c$ signal from the normalized value of 0 occur by chance (i.e. because of noise only - see \citealt{don92}, but note that this reference reports the complement probability). This was done using an optimal velocity grid (with a minimum resolution of 1.8\,km\,s$^{-1}$) representing the best compromise between increasing the signal-to-noise and not smearing the Zeeman signature. Next, the longitudinal field values $B_z$ were computed from the first-order moment of the unbinned Stokes $V$ profiles \citep{mat89,don97}, with errors derived by propagating the known uncertainties for each pixel. Similar measurements were made for the diagnostic null profiles $N$. Note that the field values for the MiMeS targets (HD\,1337, HD\,35921, and HD\,209481) were already reported by \citet{gru16}.

Since the significance level (or false-alarm probability) was larger than $10^{-3}$ in each case, corresponding to clear non-detections, upper limits on the dipolar field values were derived using Monte-Carlo simulations as described by \citet{nei15} and \citet{bla15}. For both the primary and secondary, we simulated 1000 oblique dipole models for various values of the polar magnetic field strength with random inclination angles $i$, obliquity angles $\beta$, and rotational phases. White Gaussian noise with a null average and a variance corresponding to the signal-to-noise of each observed profile was added. Using the best-fit disentangled LSD $I$ profiles, we calculated local Stokes $V$ profiles assuming the weak-field case, which were then integrated over the visible hemisphere of the star. The derived synthetic Stokes $V$ profiles were normalized to the intensity of the continuum, and the Neyman-Pearson likelihood ratio test was then applied to estimate the probability of detecting a dipolar oblique magnetic field (with a threshold of 10$^{-3}$ for the false alarm probability). A 90\% detection rate was required to consider that the simulated field would statistically be detected in the data. This translates into upper limits for the possible undetected dipolar field strength for each star and each observation. For HD\,152248 and HD\,209481, multiple observations were available and since none of them resulted in a magnetic detection, single-observation statistics were combined to extract a stricter upper limit on the non-detected field (see Sect. 4.2 of \citealt{nei15} for details). 

\begin{table*}
  \caption{Results from the high-resolution ESPaDOnS and Narval spectropolarimetry.  }
  \label{Bval2}
  \begin{tabular}{lccccccccl}
  \hline
ID & Exp. & Date & HJD & Obj & $V_r$  & $B_z$ & $N_z$ & $\sigma$ & up. lim. \\
   & /SNR &      & -2.45e6 & &  (km\,s$^{-1}$) & (G)  & (G) &   (G)  & (kG) \\
\hline
HD\,1337 & 4$\times$720 & 2009-10-09 & 5113.845 & prim & 178  & 46   & $-$37& 39 & 1.0\\
         &  1280        &            &          & sec  &$-$160&$-$46 &$-$134& 74 & 1.9\\
         &              &            &          &system&      & 35   & 40   & 60 & \\
HD\,25638& 8$\times$840 & 2013-11-20 & 6616.786 & prim & 193  & $-$62& $-$7 & 69 & 1.7\\
         &  1557        &            &          & sec  &$-$272&$-$112&$-$154& 241& 6.8\\ 
         &              &            &          & tert & 17   &$-$12 & $-$62& 55 & 2.0\\
         &              &            &          &system&      &$-$55 & 85   & 101& \\
HD\,35652& 8$\times$840 & 2016-02-27 & 7445.812 & prim &$-$198&$-$122& $-$42& 267& 7.1\\
         &  942         &            &          & sec  & 315  & 216  & 120  & 205& 4.8\\ 
         &              &            &          & tert & 124  & 189  &$-$368& 329& 8.7\\
         &              &            &          &system&      & 37   &$-$161& 249& \\
HD\,35921& 4$\times$940 & 2011-11-13 & 5878.872 & prim & 130  &$-$102& $-$44& 75 & 1.7\\
         &  1215        &            &          & sec  &$-$287& 76   & $-$48& 104& 2.8\\
         &              &            &          & tert &$-$18 & 16   & $-$45& 74 & 2.0\\
         &              &            &          &system&      & 74   &$-$15 & 98 & \\
HD\,57060&12$\times$540 & 2013-03-03 & 6354.835 & prim & 160  &  17  & $-$5 & 15 & 0.6\\
         &  2942        &            &          & sec  &$-$141& 173  & 286  &196 & 3.3\\
         &              &            &          &system&      &  92  &  102 & 47 & \\
HD\,149404& 16$\times$390& 2014-06-08 & 6816.897 & prim &$-$108& 14   & $-$10& 25 & 0.7\\
          &  3038        &            &          & sec  & 7    & 34   & 4    & 16 & 0.3\\
          &              &            &          &system&      & 35   & 9    & 16 & \\
HD\,152248& 4$\times$540 & 2014-04-13 & 6761.045 & prim &$-$135&$-$175& $-$72& 155& 4.1, {\it comb: 1.5}\\
          &  794         &            &          & sec  & 111  &416   &$-$148& 206& 5.4, {\it comb: 1.9}\\
          &              &            &          &system&      & $-$9 &$-$39 & 148& \\
HD\,152248&12$\times$540 & 2014-04-14 & 6762.060 & prim & 129  & 112  & 88   & 89 & 2.2\\
          &  1358        &            &          & sec  &$-$192&$-$153&$-$135& 104& 2.5\\
          &              &            &          &system&      &$-$46 & 7    & 97 & \\
HD\,190967& 8$\times$840 & 2016-05-18 & 7527.097 & prim &$-$32 & $-$15& $-$15& 87 & 2.1\\
          &  989         &            &          & sec  & 80   & $-$3 & 21   & 32 & 0.9\\
          &              &            &          &system&      & $-$14& $-$13& 46 & \\
HD\,209481& 4$\times$600 & 2006-12-14 & 4084.274 & prim & 74   & 14   & 60   & 74 & 1.7, {\it comb: 0.4}\\
          &  867         &            &          & sec  &$-$226& $-$36& 51   & 120& 3.1, {\it comb: 0.6}\\
          &              &            &          &system&      & $-$34& 107  & 91 & \\
HD\,209481& 4$\times$675 & 2009-07-20 & 5033.483 & prim & 27   & $-$73& 124  & 97 & 2.3\\
          &  667         &            &          & sec  &$-$100& 160  &$-$168& 134& 3.4\\
          &              &            &          &system&      & $-$46&$-$10 & 65 & \\
HD\,209481& 4$\times$825 & 2009-07-24 & 5037.485 & prim &$-$98 & 85   & $-$32& 48 & 1.1\\
          &  1191        &            &          & sec  & 212  & 213  & 24   & 83 & 2.1\\
          &              &            &          &system&      &  99  &$-$87 & 63 & \\
HD\,209481& 4$\times$675 & 2009-07-25 & 5038.492 & prim & 26   & 59   & 2    & 66 & 1.5\\
          &  916         &            &          & sec  &$-$95 & $-$51& 60   & 107& 2.7\\
          &              &            &          &system&      & $-$34& 33   & 45 & \\
HD\,209481& 4$\times$675 & 2009-07-26 & 5039.492 & prim & 46   & 192  & 67   & 62 & 1.4\\
          &  804         &            &          & sec  &$-$154&$-$225& 9    & 90 & 2.3\\
          &              &            &          &system&      & 16   & 0    & 54 & \\
HD\,209481& 4$\times$675 & 2009-07-27 & 5040.490 & prim &$-$97 & 142  &$-$112& 180& 4.2\\
          &  315         &            &          & sec  & 212  &$-$459&$-$307& 322& 8.3\\
          &              &            &          &system&      & $-$12&$-$80 & 238 & \\
HD\,209481& 4$\times$675 & 2009-07-28 & 5041.480 & prim & 10   & 26   & $-$9 & 54 & 1.2\\
          &  1058        &            &          & sec  &$-$66 & 5    & 1    & 76 & 1.9\\
          &              &            &          &system&      & $-$19& $-$2 & 29 & \\
HD\,209481& 4$\times$675 & 2009-07-29 & 5042.496 & prim & 54   & 23   & $-$62& 48 & 1.1\\
          &  1199        &            &          & sec  &$-$174& $-$21& 128  & 70 & 1.8\\
          &              &            &          &system&      & 26   & $-$19& 46 & \\
HD\,209481& 4$\times$675 & 2009-07-30 & 5043.476 & prim &$-$96 & 85   & $-$17& 60 & 1.4\\
          &  1037        &            &          & sec  & 201  & $-$62& $-$49& 100& 2.5\\
          &              &            &          &system&      & 76   & 25   & 69 & \\
HD\,209481& 4$\times$675 & 2009-07-31 & 5044.496 & prim &$-$4  & 2    & $-$39& 57 & 1.3\\
          &  1105        &            &          & sec  &$-$40 & 40   & $-$17& 86 & 2.2\\
          &              &            &          &system&      & $-$19& $-$32& 30 & \\
\hline
\end{tabular}
\end{table*}
\setcounter{table}{2}
\begin{table*}
  \caption{Continued. }
  \begin{tabular}{lccccccccl}
  \hline
ID & Exp. & Date & HJD & Obj & $V_r$  & $B_z$ & $N_z$ & $\sigma$ & up. lim. \\
   & /SNR &      & -2.45e6 & &  (km\,s$^{-1}$) & (G)  & (G) &   (G)  & (kG) \\
\hline
HD\,209481& 4$\times$675 & 2009-08-03 & 5047.461 & prim &$-$24 & 39   & $-$69& 69 & 1.6\\
          &  968         &            &          & sec  & 6    & 71   & 5    & 89 & 2.2\\
          &              &            &          &system&      & 46   & $-$2 & 32 & \\
HD\,209481& 12$\times$400& 2011-06-22 & 5735.027 & prim &$-$60 & 3    & 2    & 22 & 0.5\\
          &  2417        &            &          & sec  & 108  & $-$3 & $-$6 & 39 & 1.0\\
          &              &            &          &system&      & 1    & $-$16& 17 & \\
HD\,228854& 8$\times$840 & 2016-05-17 & 7526.067 & prim &$-$206&$-$142&$-$146& 188& 5.7\\
          &  975         &            &          & sec  &   275&$-$122& 142  & 256& 7.6\\
          &              &            &          &system&      & 27   & 208  & 269& \\
XZ\,Cep   & 8$\times$840 & 2013-08-19 & 6523.930 & prim &$-$265& $-$30& 5    & 45 & 1.1\\
          &  609         &            &          & sec  & 103  & 621  &$-$473& 488& 11.1\\
          &              &            &          &system&      & 37   &$-$180& 184& \\
\hline
\end{tabular}

{\footnotesize From left to right, the columns provide the binaries' names, the exposure lengths and peak signal-to-noise ratios in the 5000--6500\AA\ range, the observing times (both in YYYY-MM-DD format and in heliocentric Julian dates), the considered component, its heliocentric velocity, the magnetic field values and their associated errors ($\sigma$) derived from the system's Stokes $V$ profile ($B_z$) and from the null diagnostics profiles ($N_z$), and finally the upper limits on the dipolar field strength with 90\% chance of being detected. When several observations of the same target are available (as for HD\,152248 and HD\,209481), the upper limit on the undetected field from the combined statistics is quoted in italics in the last column of the first observation line after a ``comb'' prefix. }
\end{table*}

\section{Results and discussion}
\begin{figure}
\includegraphics[width=9cm]{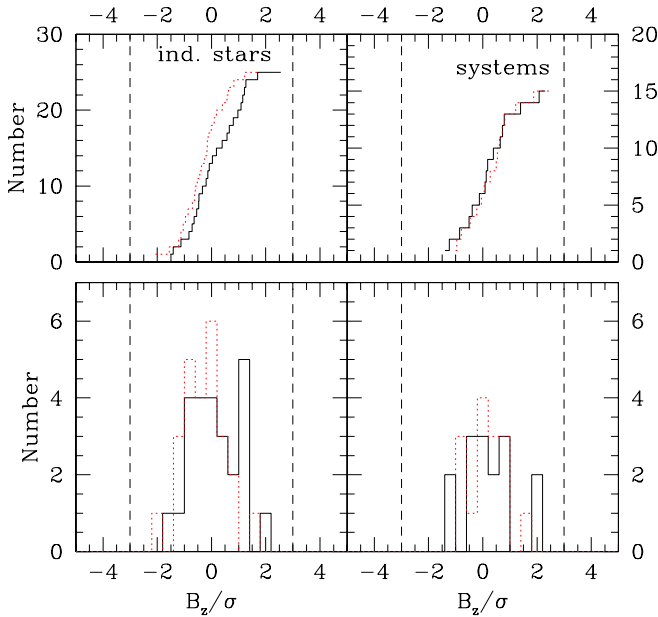}
\caption{Cumulative (top) or simple (bottom) distribution of the $B_z$ values normalized to their error bars for our sample (left, for $B_z$ values of individual stars - labelled ``primary'', ``secondary'', or ``tertiary'' in Table \ref{Bval2}; right, for $B_z$ values estimated for the whole system - from Table \ref{Bval} for FORS measurements and under ``system'' label in Table \ref{Bval2} for high-resolution observations). The dotted red lines are for the $N_z$ values, while the dashed vertical lines correspond to a 3$\sigma$ detection level. Note how the measured values are all within $2\sigma$ from zero.}
\label{histo}
\end{figure}

Table \ref{Bval} reports the resulting longitudinal magnetic field values\footnote{They are weighted averages of the contributions of all components in each system, i.e. each value is valid for the entire system. The light ratios $L_1/L_2$ of the systems are known (Table \ref{Bval}). 
 However, deriving individual limits from the system's values is not straightforward, since longitudinal fields are derived from the analysis of sets of lines, whose strengths for primary and secondary do not necessarily reflect the light ratios between components and furthermore vary with ion under consideration. We therefore refrain from deriving individual limits. } for the low-resolution spectropolarimetric data, while Table \ref{Bval2} provides the field values derived from each high-resolution spectropolarimetric observation for individual components and for each system as well as upper limits on non-detected magnetic fields per component from individual and combined observations. It is clear from these tables (see also Fig. \ref{histo}) that (1) no problem due to spurious signals affects the data (the null diagnostics being always compatible with zero within $2\sigma$ at most - and generally well within $1\sigma$) and (2) no significant magnetic field is detected for any of the systems or their components. 

One caveat must however be mentioned: when using low-resolution spectropolarimeters like FORS2, even strong fields may become unobservable, depending on the magnetic geometry of the star and the observing phase. Indeed, as the sinusoidal $B_z$ curve of magnetic oblique rotators reaches or nears a value of 0\,G at cross-over phases (i.e. when the dipole is seen equator-on), the field would become undetectable in low resolution observations. However, this occurs during a small fraction of the rotational period only, an interval depending on the error bars. For example, we may consider Plaskett's $B_z$ values \citep{gru13} and assume a sinusoidal $B_z$ varying from $-$810\,G to 680\,G. Non-detections imply $B_z\sim0$, i.e. $|B_z|<N\sigma$ with $N$ generally taken to be 3 (though a value of 5 appears better suited for FORS, see \citealt{bag02}). Since the typical error on the ten FORS measurements amounts to $\sim$55\,G, excluding the more uncertain case of LSS\,3074, non-detections would then occur during 15\% of the orbit if we consider 3$\sigma$, or 25\% for a 5$\sigma$ threshold. Missing a detection is less probable when several observations of the same target are acquired, as was done for all systems observed at low resolution. Indeed, it is unlikely to have two random exposures sampling cross-over phases for one target. Moreover, if it is already unlikely for one target, it is even more so for several independent objects. Considering a 15\% chance to miss detecting a Plaskett-like field in one observation, the probability to miss field detections in {\it all} ten observations is $(0.15)^{10}\sim10^{-8}$ (or $10^{-6}$ if 25\% is considered). Of course, there is a high chance ($\sim$80\%) to miss one detection over ten observations, but since we have two observations of each target and two targets in common with the high-resolution sample, then this risk may be mitigated. One could object that some of the observation pairs are separated by only one or two days, so that there would be no large phase change over that interval if the rotational period is long. However, it must be remembered that, contrary to the ``usual'' magnetic massive stars, our targets are relatively fast rotators in short-period systems so that $B_z$ should change rapidly, with significant variations expected on the timescale of one day, as observed for Plaskett's star. Moreover, even if we do consider that the two observations of a target were taken at the same rotational phase, the probability to miss a field in all five cases is still low: $(0.15)^{5}\sim10^{-4}$ considering a 15\% chance of missing the field detection, or $10^{-3}$ if a 25\% chance is used instead. Therefore, statistically, we could have missed detecting the magnetic field of one star in the FORS2 sample, but not more. Note that for high-resolution data, this problem does not exist as a single random observation has proven to be efficient to detect magnetism (when it exists) in a massive star during the MiMeS, BinaMIcS, and other magnetic surveys \citep[see also a more theoretical discussion by][]{pet12}. Indeed, except in very rare cases\footnote{When both components are magnetic, when the observation has been obtained at conjunction, and when the signatures of both components are of opposite shape and cancel each other either partially or totally, the magnetic signal could be diluted and not be detected if present. Note however that not only is this case very unlikely but none of our targets was observed at conjunction, as demonstrated by the very different velocities in Table \ref{Bval2} (see also Fig. \ref{highres} for a visual inspection of the components' separations).}, even if the derived longitudinal field is zero, the high-resolution Stokes $V$ profile of a magnetic object is not flat and the field can be detected whatever the phase. Missing the detection of a strong field in the high-resolution sample thus is unlikely.

It is important to note that the error bars on $B_z$ are small ($\sigma$ smaller than 100\,G), except for a few cases - for FORS data, only LSS\,3074 has large error bars ($\sigma>100$\,G) because it is a faint object ($B\sim13$\,mag compared to $B=5-9$ for the others); for ESPaDOnS/Narval data, large errors are only found for all components of HD\,35652 and HD\,228854, as well as the secondary stars of HD\,25638, HD\,57060, and XZ\,Cep. Hence our survey is generally very sensitive. In the high-resolution data, this is also reflected by the determination of the upper limits on the dipolar fields which could have remained hidden in the noise for each target. The limits are well below the field strength of Plaskett's star ($B_d>2.85$\,kG, \citealt{gru13}) for the vast majority of the cases (17 out of 25 limits on individual components). 

The objective of this paper is to test the plausibility of binary interactions as efficient processes at generating stable magnetic fields in massive systems. If this were true, we would expect all our targets to share similar properties, different from the usual massive star population. In such a case, we can further statistically characterize our sample by following the Bayesian approach of \citet[see in particular their Eq. (7), later generalised by \citealt{ase15}]{kol09}. It assumes that all stars of our sample are similar, i.e. they have a dipolar field with identical field strength, but their magnetic axes are randomly oriented with respect to the line-of-sight, and it combines the observational values into a constraint on the common field strength. We note that Eq.~(7) of \citet{kol09} is valid only for independent measurements on different stars, therefore for those objects that have been observed more than once we have considered only the field measurement with the smallest error bar. Note also that Eq.~(7) has been applied twice (see Fig. \ref{histo}): first to measurements of individual stars (results from high-resolution data for stars labelled ``primary'', ``secondary'', or ``tertiary'' in Table \ref{Bval2}), second to field estimates for the systems (combining results from Table \ref{Bval} for low-resolution observations and those listed under the ``system'' label in Table \ref{Bval2} for high-resolution observations). We then derive that there is a 90\% chance that the dipolar field strength at the pole is $B_d \le 140-220$\,G, a very stringent limit well below Plaskett's secondary field. Hence our sample and Plaskett's star, while having all underwent binary interactions, do not seem to form a specific and coherent group of magnetic stars.

To shed more light onto our results, we can compare them to the current knowledge of large-scale magnetic fields in massive stars. Such fields are generally considered to be (mostly) dipolar. In addition, except for $\zeta$\,Ori\,A (a supergiant O-star with an extremely low field, \citealt{bla15}) and $\beta$\,CMa \citep{fos15beta}, the detected fields are strong, with $B_d$ larger than 0.5\,kG for O and B0.5--1 stars \citep[e.g.][]{pet13}. This is notably the case of Plaskett's secondary. 
The detection rate of such fields, found in large samples composed of single massive stars and/or long-period massive binaries, amounts to 6--8\% \citep{fos15,gru16}. How does that compare to the case of massive interacting or post-interaction binaries? Overall, we have one magnetic detection (Plaskett's star) in 16 investigated multiple systems (our sample + Plaskett's star, comprising 35 individual stars in total). If we exclude the targets with less stringent constraints, then the high-resolution data yield one magnetic detection (Plaskett's secondary) on 19 stars. If we keep only {\it individual} measurements with field limits lower than Plaskett's field (2.85\,kG), one on 17 stars for a limit of 2\,kG, and one on 9 stars for a limit of 1\,kG. Focusing only on the mass gainers (Plaskett's secondary + those of our sample - when known, see Table \ref{prop}), there is one magnetic object out of 8 stars measured at high resolution with small error bars. In summary, whatever the case under consideration, these incidence levels appear entirely compatible with that derived from general samples, considering the small number statistics. The incidence rate in the interacting/post-interaction binary sample is thus {\it not} much larger than for the general O-star population. Therefore, there is no need to include additional processes linked to binary interactions to explain the presence of a magnetic field in Plaskett's secondary.

\section{Conclusion}
Amongst magnetic O-stars, one object is particularly remarkable: the secondary of Plaskett's star, a binary system which has very likely undergone mass-transfer in the recent past. This situation led to the intriguing possibility that binary interactions could be a source of magnetism in massive stars. To test the validity of this scenario, we have observed a sample of 15 interacting and post-interaction systems. Dedicated spectropolarimetry did not lead to any magnetic detection in these systems. In fact, for the vast majority of our targets, the longitudinal fields are $<$300\,G and the individual upper limits on the dipolar fields are below the dipolar field strength of Plaskett's secondary for 17 out of 25 binary components. Considering our targets as an homogeneous group of similar objects, a global statistical analysis of the derived field values leads to a 90\% upper limit on the (common) dipolar field strength of only 140--220\,G, underlining once again the dissimilarity between our targets and Plaskett's star. Moreover, the low rate of magnetic detection in these interacting and post-interaction systems is compatible with the rates found from general surveys of O-stars. Together with the lack of magnetic detection in Be stars, some of which are also likely products of mass-transfer, this suggests that binary interactions do not systematically trigger stable, strong magnetic fields in such systems, and that a fossil origin is still the best scenario for explaining the magnetic fields of massive stars.

\section*{Acknowledgements}
YN and GR acknowledge support from  the Fonds National de la Recherche Scientifique (Belgium), the Communaut\'e Fran\c caise de Belgique, the PRODEX XMM contract(Belspo), and an ARC grant for concerted research actions financed by the French community of Belgium (Wallonia-Brussels Federation). GAW acknowledges Discovery Grant support from the Natural Science and Engineering Research Council (NSERC) of Canada.





\begin{thebibliography}{99}
\bibitem[Alecian et al.(2015)]{ale15} Alecian, E., Neiner, C., Wade, G.~A., et al.\ 2015, New Windows on Massive Stars, 307, 330 
\bibitem[Antokhina et al.(2011)]{ant11} Antokhina, E.~A., Srinivasa Rao, M., \& Parthasarathy, M.\ 2011, \na, 16, 177 
\bibitem[Appenzeller \& Rupprecht(1992)]{app92} Appenzeller, I., \& Rupprecht, G.\ 1992, The Messenger, 67, 18 
\bibitem[Asensio Ramos et al.(2015)]{ase15} Asensio Ramos, A., Mart{\'{\i}}nez Gonz{\'a}lez, M.~J., \& Manso Sainz, R.\ 2015, \aap, 577, A125 
\bibitem[Auri{\`e}re(2003)]{aur03} Auri{\`e}re, M.\ 2003, EAS Publications Series, 9, 105 
\bibitem[Bagnulo et al.(2002)]{bag02} Bagnulo, S., Szeifert, T., Wade, G.~A., Landstreet, J.~D., \& Mathys, G.\ 2002, A\&A, 389, 191 
\bibitem[Bagnulo et al.(2009)]{bag09} Bagnulo, S., Landolfi, M., Landstreet, J.~D., et al.\ 2009, PASP, 121, 993 
\bibitem[Bagnuolo et al.(1992)]{bag92} Bagnuolo, W.~G., Jr., Gies, D.~R., \& Wiggs, M.~S.\ 1992, \apj, 385, 708 
\bibitem[Bagnuolo et al.(1994)]{bag94} Bagnuolo, W.~G., Jr., Gies, D.~R., Hahula, M.~E., Wiemker, R., \& Wiggs, M.~S.\ 1994, \apj, 423, 446 
\bibitem[Benaglia(2010)]{ben10} Benaglia, P.\ 2010, High Energy Phenomena in Massive Stars, 422, 111 
\bibitem[Blaz{\`e}re et al.(2015)]{bla15} Blaz{\`e}re, A., Neiner, C., Tkachenko, A., Bouret, J.-C., \& Rivinius, T.\ 2015, \aap, 582, A110 
\bibitem[Braithwaite(2006)]{bra06} Braithwaite, J.\ 2006, \aap, 449, 451 
\bibitem[Carrier et al.(2002)]{car02} Carrier, F., North, P., Udry, S., \& Babel, J.\ 2002, \aap, 394, 151 
\bibitem[De{\v g}irmenci et al.(1999)]{deg99} De{\v g}irmenci, {\"O}.~L., Sezer, C., Demircan, O., et al.\ 1999, \aaps, 134, 327 
\bibitem[de Mink et al.(2013)]{dem13} de Mink, S.~E., Langer, N., Izzard, R.~G., Sana, H., \& de Koter, A.\ 2013, \apj, 764, 166 
\bibitem[Djura{\v s}evi{\'c} et al.(2009)]{dju09} Djura{\v s}evi{\'c}, G., Vince, I., Khruzina, T.~S., \& Rovithis-Livaniou, E.\ 2009, \mnras, 396, 1553 
\bibitem[Donati(2003)]{don03} Donati, J.-F.\ 2003, Solar Polarization, 307, 41 
\bibitem[Donati \& Landstreet(2009)]{don09} Donati, J.-F., \& Landstreet, J.~D.\ 2009, \araa, 47, 333 
\bibitem[Donati et al.(1992)]{don92} Donati, J.-F., Semel, M., \& Rees, D.~E.\ 1992, \aap, 265, 669 
\bibitem[Donati et al.(1997)]{don97} Donati, J.-F., Semel, M., Carter, B.~D., Rees, D.~E., \& Collier Cameron, A.\ 1997, MNRAS, 291, 658 
\bibitem[Donati et al.(2002)]{don02} Donati, J.-F., Babel, J., Harries, T.~J., et al.\ 2002, MNRAS, 333, 55 
\bibitem[Donati et al.(2006)]{don06} Donati, J.-F., Howarth, I.~D., Jardine, M.~M., et al.\ 2006, MNRAS, 370, 629 
\bibitem[Drechsel et al.(1989)]{dre89} Drechsel, H., Lorenz, R., \& Mayer, P.\ 1989, \aap, 221, 49 
\bibitem[Ferrario et al.(2009)]{fer09} Ferrario, L., Pringle, J.~E., Tout, C.~A., \& Wickramasinghe, D.~T.\ 2009, \mnras, 400, L71 
\bibitem[Fossati et al.(2015a)]{fos15beta} Fossati, L., Castro, N., Morel, T., et al.\ 2015a, \aap, 574, A20 
\bibitem[Fossati et al.(2015b)]{fos15} Fossati, L., Castro, N., Sch{\"o}ller, M., et al.\ 2015b, \aap, 582, A45 
\bibitem[Gies(2003)]{gie03} Gies, D.~R.\ 2003, A Massive Star Odyssey: From Main Sequence to Supernova, 212, 91 
\bibitem[Gorda(2015)]{gor15} Gorda, S.~Y.\ 2015, Astronomy Letters, 41, 276 
\bibitem[Grunhut et al.(2013)]{gru13} Grunhut, J.~H., Wade, G.~A., Leutenegger, M., et al.\ 2013, \mnras, 428, 1686 
\bibitem[Grunhut et al.(2016)]{gru16} Grunhut, J.~H.,  Wade, G.~A., Neiner, C., et al.\ 2016, \mnras, in press (arxiv: 1610.07895)
\bibitem[Haefner et al.(1994)]{haf94} Haefner, R., Simon, K.~P., \& Fiedler, A.\ 1994, Information Bulletin on Variable Stars, 3969, 1 
\bibitem[Harries et al.(1997)]{har97} Harries, T.~J., Hilditch, R.~W., \& Hill, G.\ 1997, \mnras, 285, 277 
\bibitem[Harries et al.(1998)]{har98} Harries, T.~J., Hilditch, R.~W., \& Hill, G.\ 1998, \mnras, 295, 386 
\bibitem[Howarth et al.(1997)]{how97} Howarth, I.~D., Siebert, K.~W., Hussain, G.~A.~J., \& Prinja, R.~K.\ 1997, \mnras, 284, 265 
\bibitem[Hubrig et al.(2008)]{hub08} Hubrig, S., Sch{\"o}ller, M., Schnerr, R.~S., et al.\ 2008, \aap, 490, 793 
\bibitem[Hubrig et al.(2012)]{hub12} Hubrig, S., Kholtygin, A., Scholler, M., et al.\ 2012, Information Bulletin on Variable Stars, 6019, 1 
\bibitem[Kochukhov et al.(2010)]{koc10} Kochukhov, O., Makaganiuk, V., \& Piskunov, N.\ 2010, \aap, 524, A5 
\bibitem[Kolenberg \& Bagnulo(2009)]{kol09} Kolenberg, K., \& Bagnulo, S.\ 2009, \aap, 498, 543 
\bibitem[Kumsiashvili et al.(2005)]{kum05} Kumsiashvili, M.~I., Kochiashvili, N.~T., \& Djurasevi, G.\ 2005, Astrophysics, 48, 44 
\bibitem[Kupka et al.(1999)]{kup99} Kupka, F., Piskunov, N., Ryabchikova, T.~A., Stempels, H.~C., \& Weiss, W.~W.\ 1999, \aaps, 138, 119 
\bibitem[Langer(2014)]{lan14} Langer, N.\ 2014, Magnetic Fields throughout Stellar Evolution, 302, 1 
\bibitem[Li \& Leung(1985)]{li85} Li, Y.-F., \& Leung, K.-C.\ 1985, \apj, 298, 345 
\bibitem[Linder et al.(2006)]{lin06} Linder, N., Rauw, G., Pollock, A.~M.~T., \& Stevens, I.~R.\ 2006, \mnras, 370, 1623 
\bibitem[Linder et al.(2007)]{lin07} Linder, N., Rauw, G., Sana, H., De Becker, M., \& Gosset, E.\ 2007, \aap, 474, 193 
\bibitem[Linder (2008)]{lin08th} Linder, N., 2008, PhD thesis, Univ. of Li\`ege 
\bibitem[Linder et al.(2008)]{lin08} Linder, N., Rauw, G., Martins, F., et al.\ 2008, \aap, 489, 713 
\bibitem[Lorenz et al.(1998)]{lor98} Lorenz, R., Mayer, P., \& Drechsel, H.\ 1998, \aap, 332, 909 
\bibitem[Lorenz et al.(1994)]{lor94} Lorenz, R., Mayer, P., \& Drechsel, H.\ 1994, \aap, 291, 185 
\bibitem[McSwain \& Gies(2005)]{swa05} McSwain, M.~V., \& Gies, D.~R.\ 2005, \apjs, 161, 118 
\bibitem[Mahy et al.(2011)]{mah11} Mahy, L., Martins, F., Machado, C., Donati, J.-F., \& Bouret, J.-C.\ 2011, \aap, 533, A9 
\bibitem[Martins et al.(2010)]{mar10} Martins, F., Donati, J.-F., Marcolino, W.~L.~F., et al.\ 2010, \mnras, 407, 1423 
\bibitem[Mathys(1989)]{mat89} Mathys, G.\ 1989, \fcp, 13, 143 
\bibitem[Mayer et al.(2002)]{may02} Mayer, P., Lorenz, R., \& Drechsel, H.\ 2002, \aap, 388, 268 
\bibitem[Mayer et al.(2008)]{may08} Mayer, P., Harmanec, P., Nesslinger, S., et al.\ 2008, \aap, 481, 183 
\bibitem[Mayer et al.(2013)]{may13} Mayer, P., Drechsel, H., Harmanec, P., Yang, S., \& {\v S}lechta, M.\ 2013, \aap, 559, A22 
\bibitem[Moss(2001)]{mos01} Moss, D.\ 2001, Magnetic Fields Across the Hertzsprung-Russell Diagram, 248, 305 
\bibitem[Naz{\'e} et al.(2012)]{naz12carina} Naz{\'e}, Y., Bagnulo, S., Petit, V., et al.\ 2012, MNRAS, 423, 3413 
\bibitem[Neiner et al.(2015a)]{nei15} Neiner, C., Grunhut, J., Leroy, B., De Becker, M., \& Rauw, G.\ 2015a, \aap, 575, A66 
\bibitem[Neiner et al.(2015b)]{nei15iaus} Neiner, C., Mathis, S., Alecian, E., et al.\ 2015b, Polarimetry, proc. of IAU symposium \#305, 61 
\bibitem[Niemela et al.(1992)]{nie92} Niemela, V.~S., Cerruti, M.~A., Morrell, N.~I., \& Luna, H.~G.\ 1992, Evolutionary Processes in Interacting Binary Stars, 151, 505 
\bibitem[{\"O}zdemir et al.(2003)]{ozd03} {\"O}zdemir, S., Mayer, P., Drechsel, H., Demircan, O., \& Ak, H.\ 2003, \aap, 403, 675 
\bibitem[Palate et al.(2013)]{pal13} Palate, M., Rauw, G., Koenigsberger, G., \& Moreno, E.\ 2013, \aap, 552, A39 
\bibitem[Palate et al.(2013)]{pal13b} Palate, M., Rauw, G., \& Mahy, L.\ 2013, Central European Astrophysical Bulletin, 37, 311 
\bibitem[Palate \& Rauw(2014)]{pal14} Palate, M., \& Rauw, G.\ 2014, \aap, 572, A16 
\bibitem[Penny et al.(2002)]{pen02} Penny, L.~R., Gies, D.~R., Wise, J.~H., Stickland, D.~J., \& Lloyd, C.\ 2002, \apj, 575, 1050 
\bibitem[Penny et al.(2008)]{pen08} Penny, L.~R., Ouzts, C., \& Gies, D.~R.\ 2008, \apj, 681, 554 
\bibitem[Petit \& Wade(2012)]{pet12} Petit, V., \& Wade, G.~A.\ 2012, \mnras, 420, 773 
\bibitem[Petit et al.(2013)]{pet13} Petit, V., Owocki, S.~P., Wade, G.~A., et al.\ 2013, MNRAS, 429, 398 
\bibitem[Piskunov et al.(1995)]{pis95} Piskunov, N.~E., Kupka, F., Ryabchikova, T.~A., Weiss, W.~W., \& Jeffery, C.~S.\ 1995, \aaps, 112, 525 
\bibitem[Plaskett(1922)]{pla22} Plaskett, J.~S.\ 1922, \jrasc, 16, 284 
\bibitem[Qian et al.(2007)]{qia07} Qian, S.-B., Yuan, J.-Z., Liu, L., et al.\ 2007, \mnras, 380, 1599 
\bibitem[Rauw et al.(2001)]{rau01} Rauw, G., Naz{\'e}, Y., Carrier, F., et al.\ 2001, \aap, 368, 212 
\bibitem[Raucq et al. (2016)]{rau16} Raucq, F., Rauw, G., Gosset, E., et al.\ 2016, \aap, 588, A10 
\bibitem[Rivinius et al.(2013)]{riv13} Rivinius, T., Carciofi, A.~C., \& Martayan, C.\ 2013, \aapr, 21, 69 
\bibitem[Sana et al.(2001)]{san01} Sana, H., Rauw, G., \& Gosset, E.\ 2001, \aap, 370, 121 
\bibitem[Sana et al.(2004)]{san04} Sana, H., Stevens, I.~R., Gosset, E., Rauw, G., \& Vreux, J.-M.\ 2004, \mnras, 350, 809 
\bibitem[Schneider et al.(2016)]{sch16} Schneider, F.~R.~N., Podsiadlowski, P., Langer, N., Castro, N., \& Fossati, L.\ 2016, \mnras, 457, 2355 
\bibitem[Spruit(2002)]{spr02} Spruit, H.~C.\ 2002, \aap, 381, 923 
\bibitem[Stahl et al.(1996)]{sta96} Stahl, O., Kaufer, A., Rivinius, T., et al.\ 1996, \aap, 312, 539 
\bibitem[Stickland et al.(1994)]{sti94} Stickland, D.~J., Koch, R.~H., Pachoulakis, I.,  \& Pfeiffer, R.~J.\ 1994, The Observatory, 114, 107 
\bibitem[Stickland(1997)]{sti97} Stickland, D.~J.\ 1997, The Observatory, 117, 37 
\bibitem[Szeifert et al.(1998)]{sze98} Szeifert, T., Appenzeller, I., Fuertig, W., et al.\ 1998, \procspie, 3355, 20 
\bibitem[Tamajo et al.(2012)]{tam12} Tamajo, E., Munari, U., Siviero, A., Tomasella, L., \& Dallaporta, S.\ 2012, \aap, 539, A139 
\bibitem[Tout et al.(2008)]{tou08} Tout, C.~A., Wickramasinghe, D.~T., Liebert, J., Ferrario, L., \& Pringle, J.~E.\ 2008, \mnras, 387, 897 
\bibitem[Thaller et al.(2001)]{tha01} Thaller, M.~L., Gies, D.~R., Fullerton, A.~W., Kaper, L., \& Wiemker, R.\ 2001, \apj, 554, 1070 
\bibitem[van Bever \& Vanbeveren(1997)]{van97} van Bever, J., \& Vanbeveren, D.\ 1997, \aap, 322, 116 
\bibitem[Wade et al.(2011)]{wad11} Wade, G.~A., Howarth, I.~D., Townsend, R.~H.~D., et al.\ 2011, \mnras, 416, 3160 
\bibitem[Wade et al.(2012a)]{wad12} Wade, G.~A., Grunhut, J., Gr{\"a}fener, G., et al.\ 2012a, \mnras, 419, 2459 
\bibitem[Wade et al.(2012b)]{wad12dash} Wade, G.~A., Ma{\'{\i}}z Apell{\'a}niz, J., Martins, F., et al.\ 2012b, \mnras, 425, 1278 
\bibitem[Wade et al.(2015)]{wad15} Wade, G.~A., Barb{\'a}, R.~H., Grunhut, J., et al.\ 2015, \mnras, 447, 2551 
\bibitem[Wade et al.(2016a)]{wad16} Wade, G.~A., Neiner, C., Alecian, E., et al.\ 2016a, \mnras, 456, 2 
\bibitem[Wade et al.(2016b)]{wad16be} Wade, G.~A., Petit, V., Grunhut, J.~H., Neiner, C., \& MiMeS Collaboration 2016b, Bright Emissaries: Be Stars as Messengers of Star-Disk Physics, 506, 207 
\bibitem[Wickramasinghe et al.(2014)]{wic14} Wickramasinghe, D.~T., Tout, C.~A., \& Ferrario, L.\ 2014, \mnras, 437, 675 
\bibitem[Ya{\c s}arsoy \& Yakut(2013)]{yas13} Ya{\c s}arsoy, B., \& Yakut, K.\ 2013, \aj, 145, 9 
\bibitem[Zhao et al.(2014)]{zha14} Zhao, E., Qian, S., Li, L., et al.\ 2014, \na, 26, 112 
\end{thebibliography}


\bsp	
\label{lastpage}
\end{document}